\documentclass[5p]{elsarticle}

\usepackage[draft]{hyperref}
\usepackage{graphicx}
\usepackage{subfig}
\usepackage{aasmacros}
\usepackage{outlines}
\usepackage{xcolor}

\usepackage{array, booktabs, makecell}
\usepackage[version=4]{mhchem}
\usepackage{siunitx}

\journal{Icarus}

\usepackage{numcompress}\biboptions{authoryear}

\begin{document}

\begin{frontmatter}

\title{Dynamics of Small Bodies in Orbits Between Jupiter and Saturn}

\author[mymainaddress]{Andrew C. Roberts\corref{mycorrespondingauthor}}
\cortext[mycorrespondingauthor]{Corresponding author. Current Address: Department of Physics, Carnegie Mellon University, 5000 Forbes Avenue, Pittsburgh, PA 15213, 412-268-2740}
\ead{acrobert@andrew.cmu.edu}

\author[mysecondaryaddress]{Marco A. Mu\~{n}oz-Guti\'errez}

\address[mymainaddress]{University of Pennsylvania, 1600 John F Kennedy Boulevard, Philadelphia, USA}
\address[mysecondaryaddress]{Academia Sinica Institute of Astronomy and Astrophysics, 11F of AS/NTU Astronomy-Mathematics Building, No.1, Sec. 4, Roosevelt Road, Taipei 10617, Taiwan, R.O.C.}

\begin{abstract}

We examine the dynamics of small bodies in orbits similar to that of comet 29P/Schwassmann-Wachmann 1, i.e. near-circular orbits between Jupiter and Saturn. As of late 2019, there are 14 other known bodies in this region that lie in similar orbits. Previous research has shown that this region of the solar system, which in this work we call the ``near Centaur region'' (NCR), is not stable, suggesting that any bodies found in it would have very short lifetimes. We performed 20 Myr high-precision numerical simulations of the evolution of massless particles, initially located in the Kuiper belt but close to Neptune, with perihelia slightly below 33 au (``Neptune crossers''). Some of these particles quickly migrate inward, passing through the NCR before becoming Jupiter Family Comets. We find that objects in the NCR do indeed generally travel through it very quickly. However, our simulations reveal that resonant behavior in this region is somewhat common and can trap objects for up to 100 Kyr. We summarize the dynamics of 29P and other observed bodies in the region---two of which seem to be clearly exhibiting resonant behavior---and use our simulations to put limits on the reservoir size of the Neptune crossers, which at its time determines the reservoir size of the Kuiper belt. Finally, we put some constraints on the current population of Centaurs and particularly of the NCR population, based on the  injection rates required to keep the observed population of Jupiter family comets in steady-state.

\end{abstract}


\end{frontmatter}


\section{Introduction}

The existence of a comet belt beyond Neptune, now commonly referred as the Kuiper belt, was originally proposed as the source of the Short Period Comets \citep[SPCs;][]{Fernandez80}. Since then, this at first theoretical idea was strengthened, and its existence later confirmed, by further numerical simulations \citep{Duncan88,Quinn90} and by the discovery of the first object beyond Neptune \citep[apart from Pluto;][]{Jewitt93}. 

Objects in the Kuiper belt are put into trajectories strongly perturbed by Neptune (those with perihelia below $\sim33$ au) by secular and chaotic perturbations from the giant planets, as well as by encounters with other large objects in the belt, such as dwarf planets \citep{Torbett89,Duncan95,Morbidelli97,Dones15,Nesvorny17,Munoz19}. Once a cometary nucleus reach Neptune's influence region, it has a chance to be scattered inward or outward over time. If scattered inwards, until reaching the influence region of Uranus, a similar process takes place, so the objects can be handed down by the giant planets, up to the stage when they become Jupiter Family Comets (JFCs), where they are dominated by Jupiter's gravity, keeping an almost constant Tisserand parameter \citep{Levison97}. 

JFCs are dynamically short-lived, and they can end up being ejected from the solar system, colliding with the Sun or a planet. Apart from this, they can be split by tidal forces during a close encounter. The short life of the JFC population suggests a quasi-constant injection rate of new objects from the reservoir region, the Kuiper belt, under the assumption of a population in steady state \citep[see for example][for a recent review of cometary dynamics and their reservoirs]{Dones15}. 

Centaurs are thought to represent the intermediate stage between Kuiper Belt Objects (KBOs) and JFCs. In this work, we define Centaurs \citep[as in][]{Jewitt90} as objects that satisfy two criteria: 
\begin{itemize}
\item The object's perihelion, $q$, and semimajor axis, $a$, satisfy $a_J < q < a_N$ and $a_J < a < a_N$, where $a_J$ and $a_N$ are the semimajor axes of Jupiter and Neptune, at $5.204$ au and $30.11$ au, respectively.
\item The object is not permanently captured in a 1:1 mean motion resonance (MMR) with any of the giant planets, i.e. it is not a Trojan, though it may briefly reside in such a resonance. 
\end{itemize}
There are several interesting characteristics of Centaurs as they migrate through the outer solar system. Because of the processes by which these objects migrate, they are almost always crossing the orbit of at least one of the giant planets. For example, the aphelion of an object evolving primarily under the influence of Uranus can be pulled inward until the object's perihelion crosses Saturn's orbit, when its aphelion can then be brought inward. It is in general relatively unlikely that the object's orbit circularizes in between two planets, though this can happen if it gets caught in a resonance as is discussed later on in this work. This sort of evolution means that the object's Tisserand parameter, $T_P = a_P/a + 2 \cos{i} \sqrt{a/a_P (1-e^2)}$, with respect to a planet, is almost always slightly below 3 when it is crossing that planet's orbit\footnote{Also, notably, objects with Tisserand parameters closer to 3 are more likely to be scattered inward, while a lower $T_P$ near 2 makes outward scattering and/or ejection more likely \citep{Levison97,DiSisto09}}, where $a_P$ is the planet's semimajor axis and $a$, $e$, and $i$ are the object's semimajor axis, eccentricity, and inclination \citep{Levison97}.

To date, over 400 centaurs have been discovered, some of which are active comets \citep{Jewitt09}. Of particular interest in this work is comet 29P/Schwassmann-Wachmann 1 (29P henceforth), which is known for its frequent outbursts during which it brightens by several magnitudes. It also resides on a near-circular orbit just outside of Jupiter's, with semimajor axis $a = 5.986$ au, eccentricity $e = 0.044$, and an inclination of $I = 9.39 ^{\circ}$. Having been discovered in 1927, its long history of observation combined with its unique activity and orbit have made it the focus of much research in the time since \citep[see for example][]{Jewitt90,Stansberry04,Ivanova16}. 

29P resides in a special region right before where Centaurs typically become JFCs. In this region, Centaurs first begin to experience the influence of Jupiter, whose gravitational potential dominates the JFC dynamics. Additionally, at the distances of the order of Jupiter's and Saturn's heliocentric locations, water ice can hardly sublimate due to low temperatures, and is not sufficient to be the main source of the material that drive the activity of comets in this region. Rather, more volatile compounds are believed to be responsible for the start of the cometary activity in this regions, likely CO and N$_2$ ices, as well as phase transitions from amorphous to crystalline states \citep{Senay94,Jewitt09}.

These characteristics confer to the region between Jupiter and Saturn a special interest, since it represents a transitional zone between the outer parts of the planetary system, where Centaurs slowly evolve while remaining largely inactive, and the region where the dramatic dynamics and activity of JFCs takes place.

Perhaps the most notable feature of this dynamical region is its quick evolution. Previous research has shown that the region between Jupiter and Saturn is unstable, such that orbits greatly change on very short timescales \citep[e.g.][]{DiSisto20}. In particular, a frequency map analysis \citep[FMA;][]{Laskar90,Laskar92,Laskar93}, performed for a broad range of distances in the solar system, showed that there are no long-life or stable regions for orbits with semimajor axes between 5 and 10 au, with the exception of the Jupiter Trojans \citep{Robutel01}. 

In this work, we examine the dynamics of the Near Centaurs (NCs), which we define to be objects between Jupiter and Saturn whose orbits do not cross the semimajor axis of either planet and are not Jupiter Trojans; this is, they have perihelion $q > 5.204$ au and aphelion $5.6 < Q < 9.583$ au. We have found 15 known objects in such orbits, including 29P, by using the JPL Small Body Database Search Engine \footnote{\url{https://ssd.jpl.nasa.gov/sbdb_query.cgi}}. 

\begin{figure}[!ht]
\centering
\includegraphics[width=\linewidth]{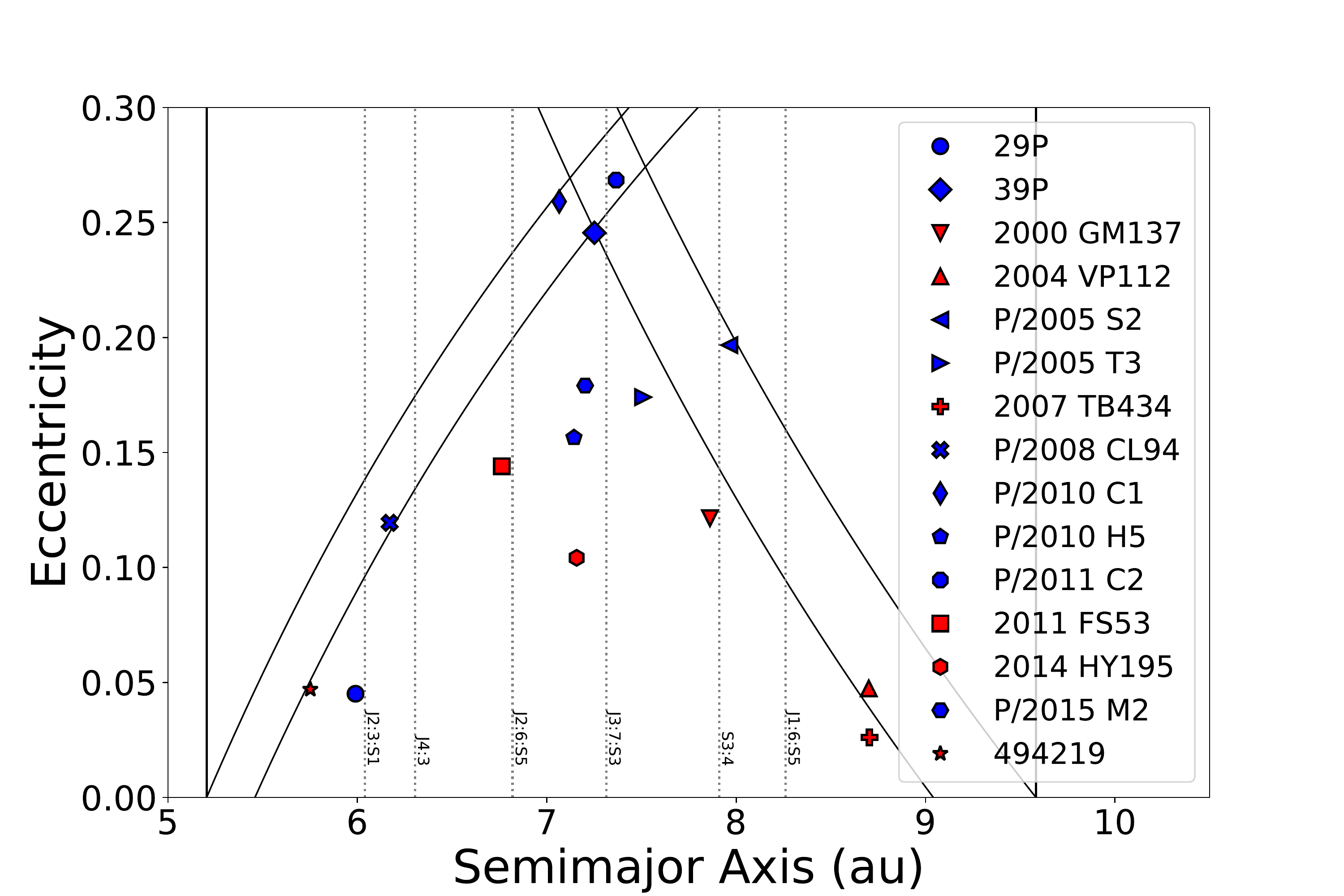}
\caption{Scatterplot of current NCR objects, active in blue and inactive in red. Several resonances are labeled with vertical dashed lines. \label{fig:realNCs}}
\end{figure}

In our sample, nine objects are active comets; additionally, four have $e < 0.05$. The orbital characteristics of these bodies are listed in Table \ref{tab:realNCs} and plotted in Figure \ref{fig:realNCs}. If these bodies were in fact brought to their current location by our understanding of how Centaurs evolve, it is somewhat surprising to find so many. Studying the histories of these objects and the dynamics of the NC region will shed light on our understanding of cometary migration and the handoff between the Centaurs and the JFCs. We search for possible unexpected areas of stability in the NC region, by performing an updated FMA on a wide region of phase-space between the orbits of Jupiter and Saturn. We also explore the dynamics of NCs through studying the 15 known objects themselves and simulated particles originating in the Kuiper belt.

\begin{table*}[!htb]
\centering
\begin{tabular}{||c|c c c c c||} 
\hline
Name & $a$ (au) & $e$ & $i$ (deg) & $q$ (au) & $Q$ (au) \\ [0.5ex] 
\hline\hline
29P/Schwassmann-Wachmann 1 & 5.99 & 0.045 & 9.391 & 5.72 & 6.26  \\ 
\hline
39P/Oterma & 7.251 & 0.246 & 1.943 & 5.471 & 9.032  \\ 
\hline
2000 GM137 & 7.862 & 0.121 & 15.865 & 6.907 & 8.816  \\ 
\hline
2004 VP112 & 8.7 & 0.047 & 6.851 & 8.288 & 9.112  \\ 
\hline
P/2005 S2 (Skiff) & 7.965 & 0.197 & 3.141 & 6.398 & 9.531  \\ 
\hline
P/2005 T3 (Read) & 7.509 & 0.174 & 6.261 & 6.202 & 8.816  \\ 
\hline
2007 TB434 & 8.704 & 0.026 & 9.768 & 8.477 & 8.931  \\ 
\hline
P/2008 CL94 (Lemmon) & 6.171 & 0.119 & 8.348 & 5.434 & 6.907  \\ 
\hline
P/2010 C1 (Scotti) & 7.066 & 0.259 & 9.142 & 5.235 & 8.896  \\ 
\hline
P/2010 H5 (Scotti) & 7.144 & 0.156 & 14.087 & 6.026 & 8.261  \\ 
\hline
P/2011 C2 (Gibbs) & 7.366 & 0.268 & 10.911 & 5.389 & 9.344  \\ 
\hline
2011 FS53 & 6.763 & 0.144 & 7.645 & 5.789 & 7.737  \\ 
\hline
2014 HY195 & 7.158 & 0.104 & 22.896 & 6.412 & 7.903  \\ 
\hline
P/2015 M2 (PANSTARRS) & 7.203 & 0.179 & 3.974 & 5.913 & 8.492  \\ 
\hline
494219 & 5.751 & 0.047 & 43.422 & 5.481 & 6.022  \\ 
\hline
\end{tabular}
\caption{Orbital elements and activity of all NCR objects discovered to date, sorted by year of discovery.}
\label{tab:realNCs}
\end{table*}

Jupiter and Saturn are very close to a 5:2 MMR with each other \citep[for an in-depth study of this phenomenon see, e.g.][]{Michtchenko01}. This creates zones in the NCR in nearly three-body resonance (3BR) with both planets \citep{Gallardo14}. In this work we observed resonant behavior at locations inside the NCR corresponding to the overlapping of first order MMRs with both Jupiter and Saturn, i.e. resonances which result from a chain of 2 two body MMRs, also called ``non-pure'' 3BR \citep{Gallardo16}. Here we label these kind of resonances as J$k_1$:$k_2$:S$k_3$, such that integers $k_i$ express the resonant condition $k_1n_J + k_2n_p + k_3n_S\approx0$, where $n_J$, $n_S$, and $n_P$ are the mean motions of Jupiter, Saturn, and the test particle, respectively. We observed in particular Jupiter-particle-Saturn ratios such as J2:3:S1 at $6.04$ au, J2:6:S5 at $6.82$ au, J3:7:S3 at $7.315$ au, and  J1:6:S5 at $8.261$ au. We show that particles close to these locations and at sufficiently low eccentricity would avoid stronger perturbations by either of the planets, therefore they could remain in the NCR for much longer than is expected.

This paper is organized as follows: in Section 2, we describe our simulations and methods. In Section 3, we present and discuss the results of our simulations. We present an updated stability map for the region between Jupiter and Saturn based on frequency analysis; we analyze Centaur evolution and present some constraints on the population of objects larger than 2 km in diameter both for the Centaurs and for their parent populations in the Kuiper Belt. In Section 4, we show the results from the detailed short-term simulations of the fifteen known NCR objects, compare them to our simulated particles, discuss the general dynamics of the region, and estimate the number of real Near Centaurs. In Section 5, we discuss and compare our results with previous studies. Finally in Section 6, we present our conclusions.

\section{Simulations}

\begin{figure*}[!ht]
   \centering
   \subfloat[][]{\includegraphics[width=.475\textwidth]{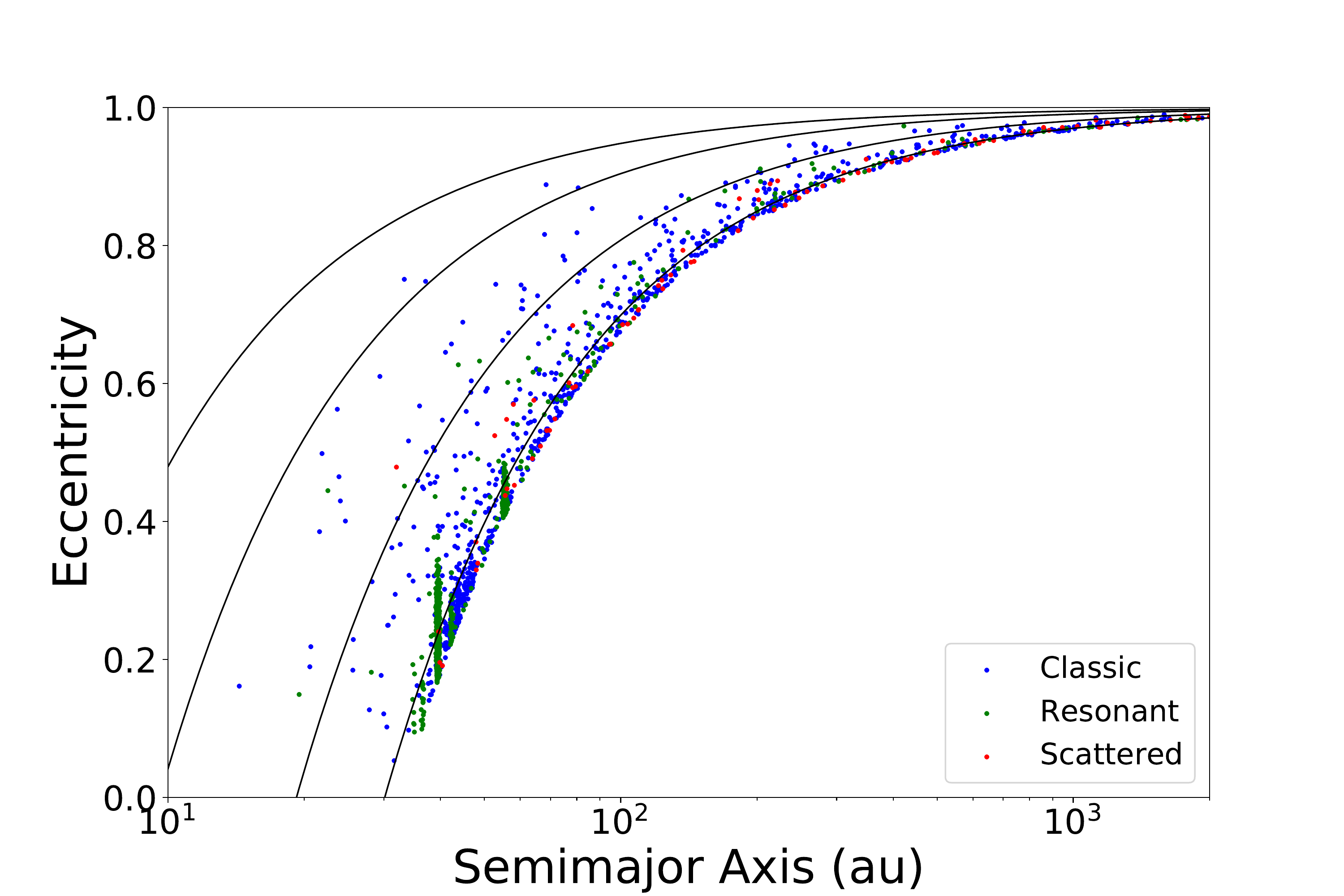}}\quad
   \subfloat[][]{\includegraphics[width=.475\textwidth]{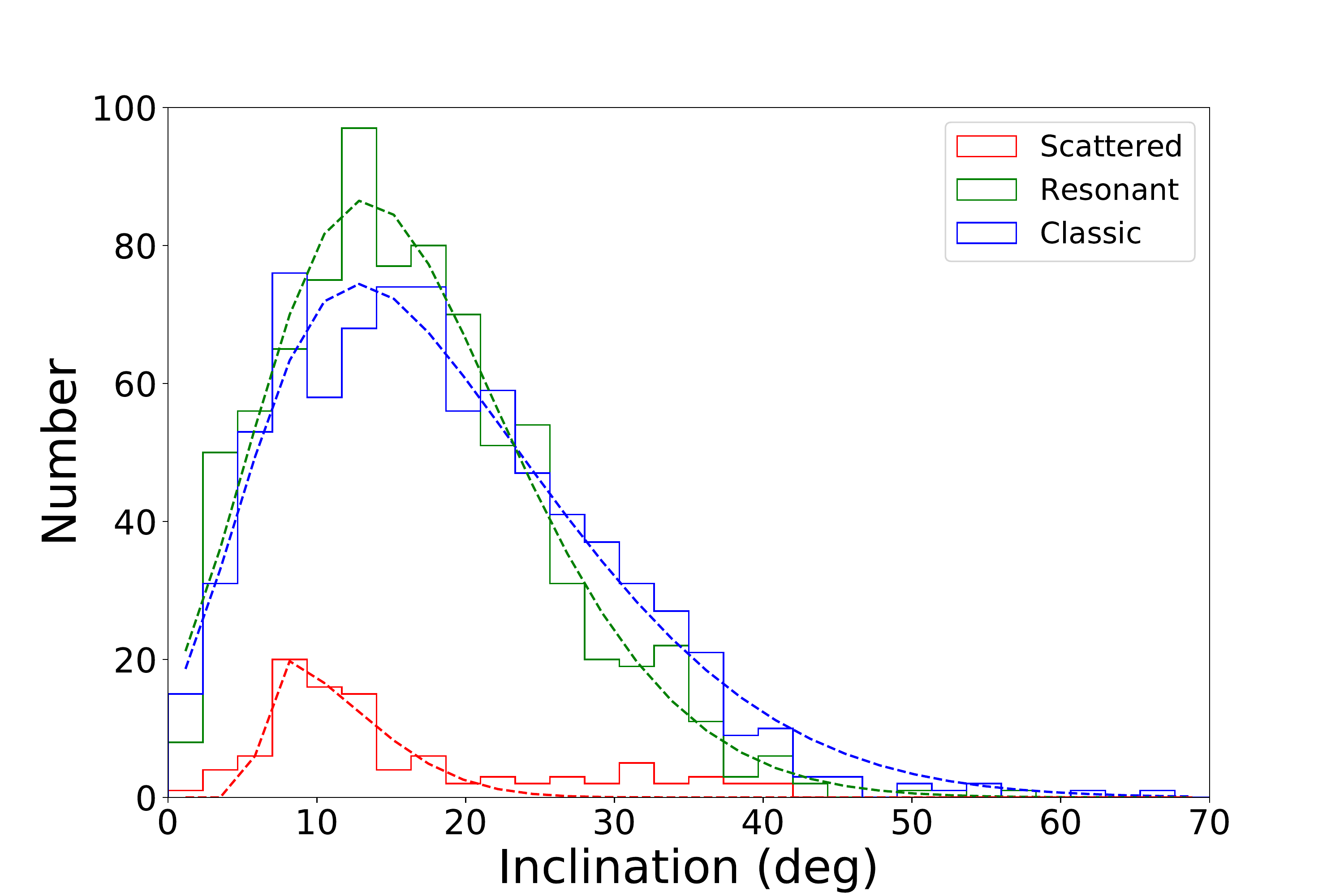}}\\
   \subfloat[][]{\includegraphics[width=.475\textwidth]{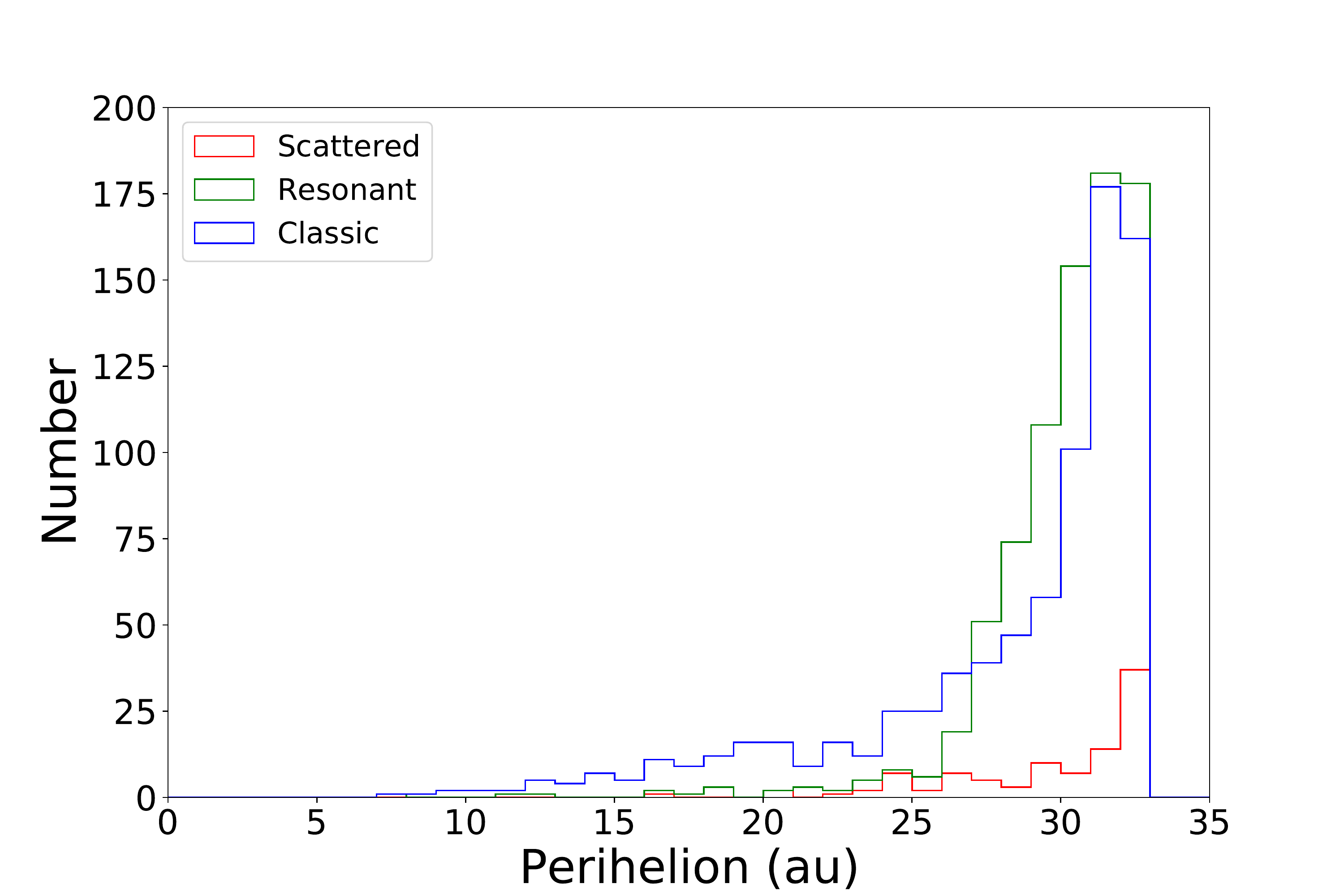}}\quad
   \subfloat[][]{\includegraphics[width=.475\textwidth]{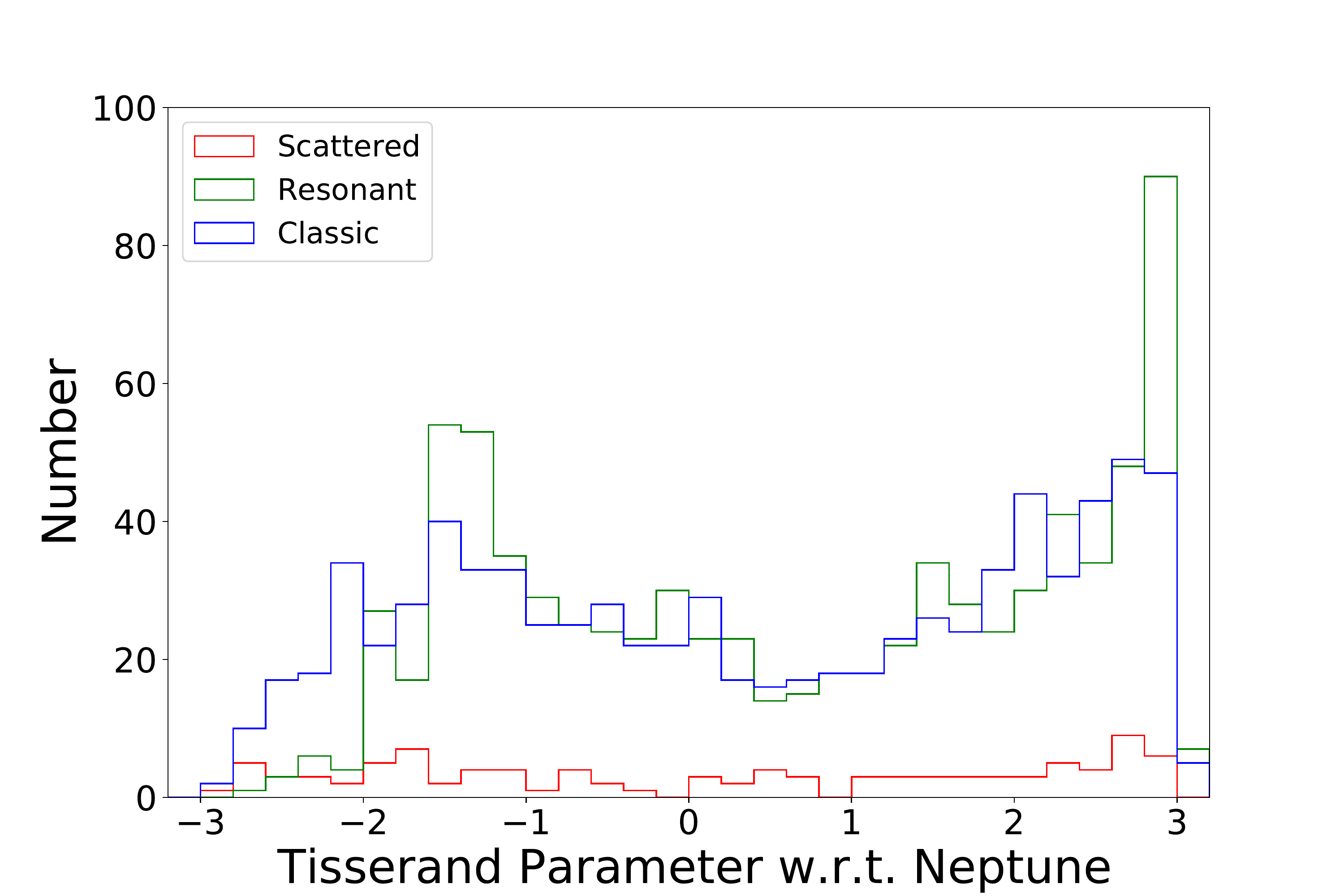}}
   \caption{A summary of the initial conditions of simulated massless particles. The colors correspond to the population of origin of each particle; all of our particles, regardless of origin, start as Neptune Crossers. \label{fig:init}}
\end{figure*}

To gain a general understanding of the dynamical behaviour and stability in the NCR, we first performed a short-term integration ($\sim 8.3\times10^4$ yr), using the B\"ulirsch-St\"oer integrator from the Mercury package \citep{Chambers99}. We integrated thousands of test particles, covering a wide region of phase space, from 4 to 12 au in semimajor axis and from 0 to 0.25 in eccentricity. We analyse the stability of the region by performing a FMA \citep{Laskar90,Laskar92,Laskar93} and calculating the diffusion parameter of each particle, as has been done elsewhere for different dynamical systems \citep[see for example][]{Robutel01,Correia05,Munoz17a}.

The rest of our simulations were performed with the public code REBOUND 3.8.1 \citep{Rein12,Rein17} using the IAS15 integrator \citep{Rein15}, which uses a dynamic time-step to accurately simulate close encounters. The simulation time-step parameter was set to a tenth of a year, though the actual time-step was generally closer to one hundredth of a year. All of our simulations were performed with seven of the eight planets; Mercury was removed and its mass added to that of the Sun to avoid the need for relativistic corrections. 

We simulated an initial sample of just under 1700 test particles, each with initial perihelia inside $2 \sqrt{3}$ times Neptune's Hill Radius, i.e. $q < 32.77$ au (we call these particles ``Neptune crossers''). These particles constitute the final distribution of crossers obtained after 1 Gyr of evolution in a simulation of the Kuiper Belt, under the influence of the giant planets and the 34 largest discovered objects in the belt itself \citep[taken from][]{Munoz19}. In the simulations of \citeauthor{Munoz19} each particle comes initially from one of three populations in the Kuiper belt: the Classical, the Resonant, and the Scattering. Of the particles simulated in this work, 800 came originally from the Classical belt, 800 from the Resonant populations, and 98 from the Scattering disk. A summary of the initial distribution of the orbits of simulated objects can be found in Figure \ref{fig:init}. The initial values of the angles, argument of perihelion, $\omega$, longitude of ascending node, $\Omega$, and mean anomaly, $M$, were assigned at random between $0$ and $2\pi$. 

The particles and planets were simulated for 20 Myr. Every 20 kiloyears, any Centaurs (under the above definition) present in the simulation (and that had not been previously cloned) were cloned five times by adjusting each of their spatial coordinates by a number chosen uniformly at random in $\pm 1\times10^{-7}$ au \citep[as in][]{Levison97}. Additionally, if any particle had a semimajor axis less than zero (hyperbolic) or greater than 20,000 au (when its period begins to approach the length of the simulation), it was removed. 

To improve the statistics from our Scattering population, we performed simulations of the Scattering particles four times, each time starting with different random initial angles and cloning centaurs ten times instead of five. This gave us an effective reservoir size of 4900 particles, comparable to that of the particles coming from the Classical and Resonant populations (4800 effective particles each).

\section{Results and Discussion}

\subsection{Global Dynamical Characterization of the Region Between Jupiter and Saturn}

\begin{figure*}
\centering
\includegraphics[width=\linewidth]{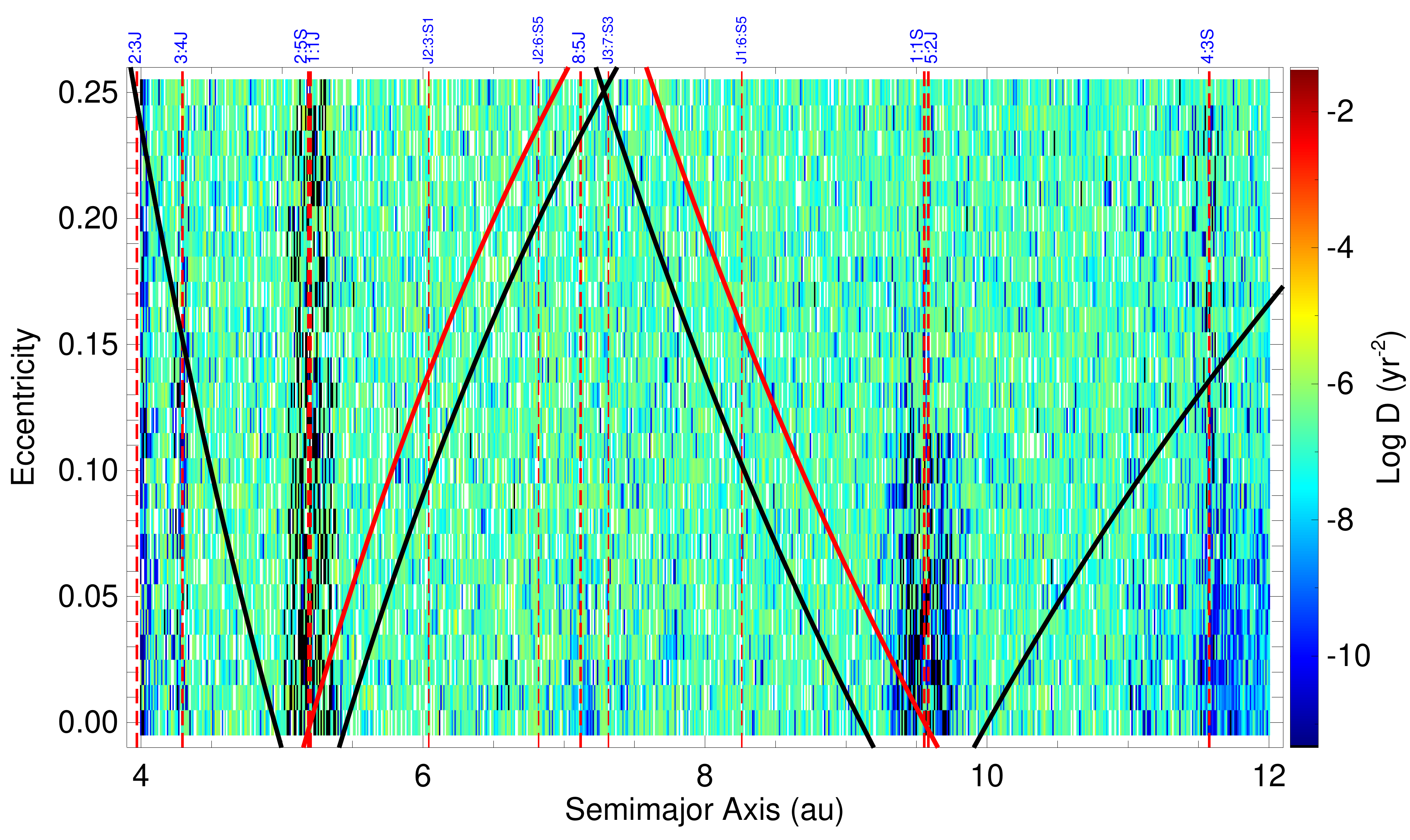}
\caption{Diffusion map of a wide region of phase-space between the orbits of Jupiter and Saturn, performed with a FMA over thousands of test-particles. The initial inclination of all particles is set to zero, while angles are started with random values between 0 and 360$^\circ$.\label{fig:FAmap}}
\end{figure*}

In a previous work, \citet{Robutel01} carried out extensive numerical simulations of the solar system, in order to identify stable and unstable zones along all of the system. \citeauthor{Robutel01} used the frequency analysis technique to obtain the main frequencies, $\nu$, of the orbits of particles in the solar system by defining a dynamical variable, $z_j'(t)$, closely related to a complex combination of the actions, $J_j$, and angles, $\theta_j$, of the system. This variable, defined as $z_j'(t)= a(t) \exp(i\lambda(t))$ (where $a$ and $\lambda$ are the semimajor axis and mean longitude of each particle, respectively) is related to the complex combination $z_j(t) = J_j \exp(i\theta_j(t))$, which is a direct function of the actions and angles of the orbit. It is shown that $z_j' = f(z_1, z_2, z_3, ..., z_n)$, where $f$ is a function close to the identity. In this way, even if $z'_j$ is not directly a function of the actions and angles, its study through frequency analysis still provides the main frequencies of the system, as long as the orbits are regular, since in that case the actions will be constants and the angles will evolve linearly with the obtained frequencies. On the contrary, for unstable orbits, the main frequencies will vary randomly and their rate of change will provide a useful measure of their chaoticity.

\citeauthor{Robutel01} used a set of initial conditions that covered from 0.38 to 100 au in semimajor axis. They found no stable regions between the orbits of Jupiter and Saturn, except for the Jupiter Trojans. Since the FMA method allows for the determination of the mean frequency, $\nu$, of an orbit, if performed at consecutive time intervals in a short-term simulation (long enough only as for the calculation of the frequencies be done with high precision, which in practice is achieved after a couple thousand orbital revolutions), it is possible to define a diffusion parameter, $D$, as \citep{Correia05,Munoz17a}:
\begin{equation}
D=\frac{\nu_1-\nu_2}{T},
\end{equation}
where $T$ is half the total length of the simulation and $\nu_1$ and $\nu_2$ are the main frequencies of the orbit, calculated on the adjacent time intervals. It is known that a quasi-periodic (regular) orbit will keep almost constant frequencies along the simulation, thus its diffusion parameter will be small; on the other hand, large changes among the main frequencies point to the overall instability of the trajectories and consequently to a larger value of the diffusion parameter.

We perform a FMA to a set of $20\,826$ test particles arranged in a grid of initial conditions covering a wide region of phase-space in semimajor axis (from 4 to 12 au) and eccentricity (from 0 to 0.25). The inclination of all particles was set to zero but the values of the angles, $\omega$, $\Omega$, and $M$ were assigned randomly between zero and 360$^\circ$. The FMA was performed  using the improved algorithm of \citet{Sidly96}.

Fig. \ref{fig:FAmap} shows the resulting diffusion map of the region from 4 to 12 au, where the NCR is shown delimited by solid red curves, while solid black curves stand for the collision lines at the location of the perihelion and aphelion of Jupiter and Saturn. It is evident that no long-term stability areas exist between the orbits of Jupiter and Saturn, except for coorbital particles, i.e. in 1:1 MMR with either of the planets. The overall diffusion time for the green-colored region, which occupies most of the map surface, is approximately $1.1\times10^5$ yr. This simple estimation was done by considering $t_{\it Diff}=1/(D'P)$, where $D'$ is the median value of the diffusion parameter for the whole map, while $P$ is an average period of 20 yr, used as a proxy value for the orbits in the region.

The apparent stability of the Saturn Trojan region is somewhat surprising, given the lack of real objects observed in that location. However, this result is not completely unexpected. In a previous work, \cite{Nesvorny02} found that for orbits with eccentricities above $\sim0.12$ (in great accordance with our results) the overlapping of the near 5:2 MMR between Jupiter and Saturn with the 1:1 MMR of Saturn's coorbital particles, creates a chaotic region where particles are not stable. Below $e=0.12$, orbital stability in short-term integrations is also found by \citeauthor{Nesvorny02}. However, in longer simulations (4 Gyr long) they found that the population of Saturn Trojans is depleted by a factor of $\sim100$, mainly due to secular perturbations that tend to increase the eccentricity of the particles. 

In our simulations, the diffusion time for hypothetical Saturnian Trojans as obtained from the diffusion map of Fig. \ref{fig:FAmap} is only $\approx4.6\sim10^5$ yr, considering the mean $D$ value for a rectangular region defined by the limits: $e<0.12$ and $a_{\it Sat}-0.7<a<a_{\it Sat}+0.7$, where $a_{\it Sat}$ is the semimajor axis of Saturn, together with a period equal to that of Saturn (29.53 yr). 

The short diffusion time scale of Saturn Trojans is in accordance with the expectation from secular perturbations that finally lead to the depletion of any primordial population \citep{Nesvorny02}. Even if Saturn were able to recapture objects in coorbital motion after the scattering event thought to be caused by the early migration of the giant planets, in a similar way as Jupiter have been shown to be able to repopulate its stable zones around the L4 and L5 Lagrange points \citep{Nesvorny13}, the lifetime of the recaptured Saturnian Trojans would be very short as to represent a significant observable population at any given time during the solar system lifetime.

In the diffusion map, we highlight MMR regions which are clearly identifiable from the FMA. The location of two-body MMRs with either Jupiter or Saturn is indicated by thick dashed red lines, while the mean motion ratios and a label indicating the involved planet (J for Jupiter, S for Saturn) are shown at the top of the Figure. We also indicate with thin dashed red lines, the location of possible three-body MMRs involving Jupiter, Saturn, and a particle, that seem to dominate the dynamics of some objects in the NCR, as we show ahead in this work. 

\subsection{Centaur Dynamics}

\begin{figure*}[!ht]
   \centering
   \subfloat[][]
   {\includegraphics[width=.475\textwidth]{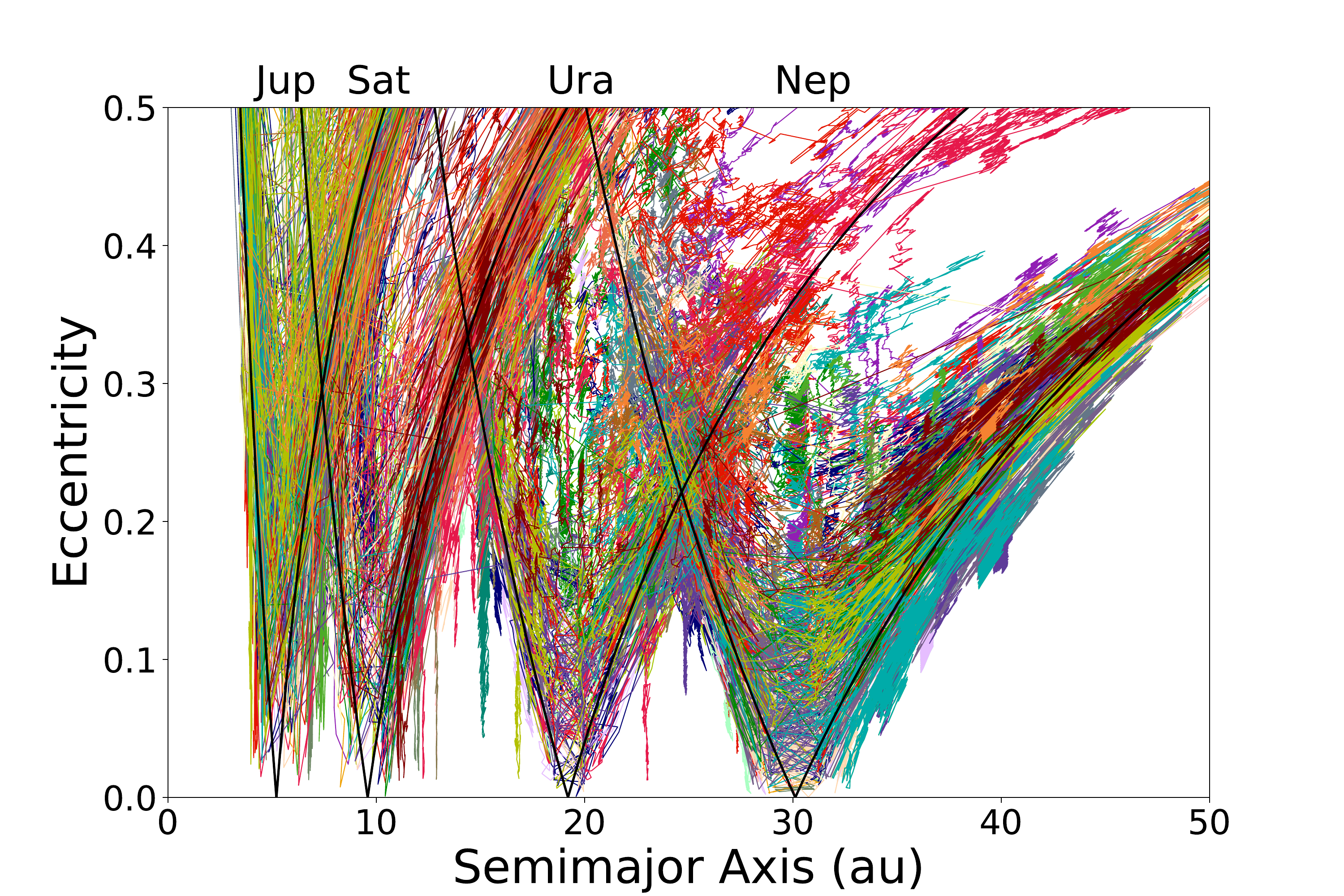}}
   \subfloat[][]
   {\includegraphics[width=.475\textwidth]{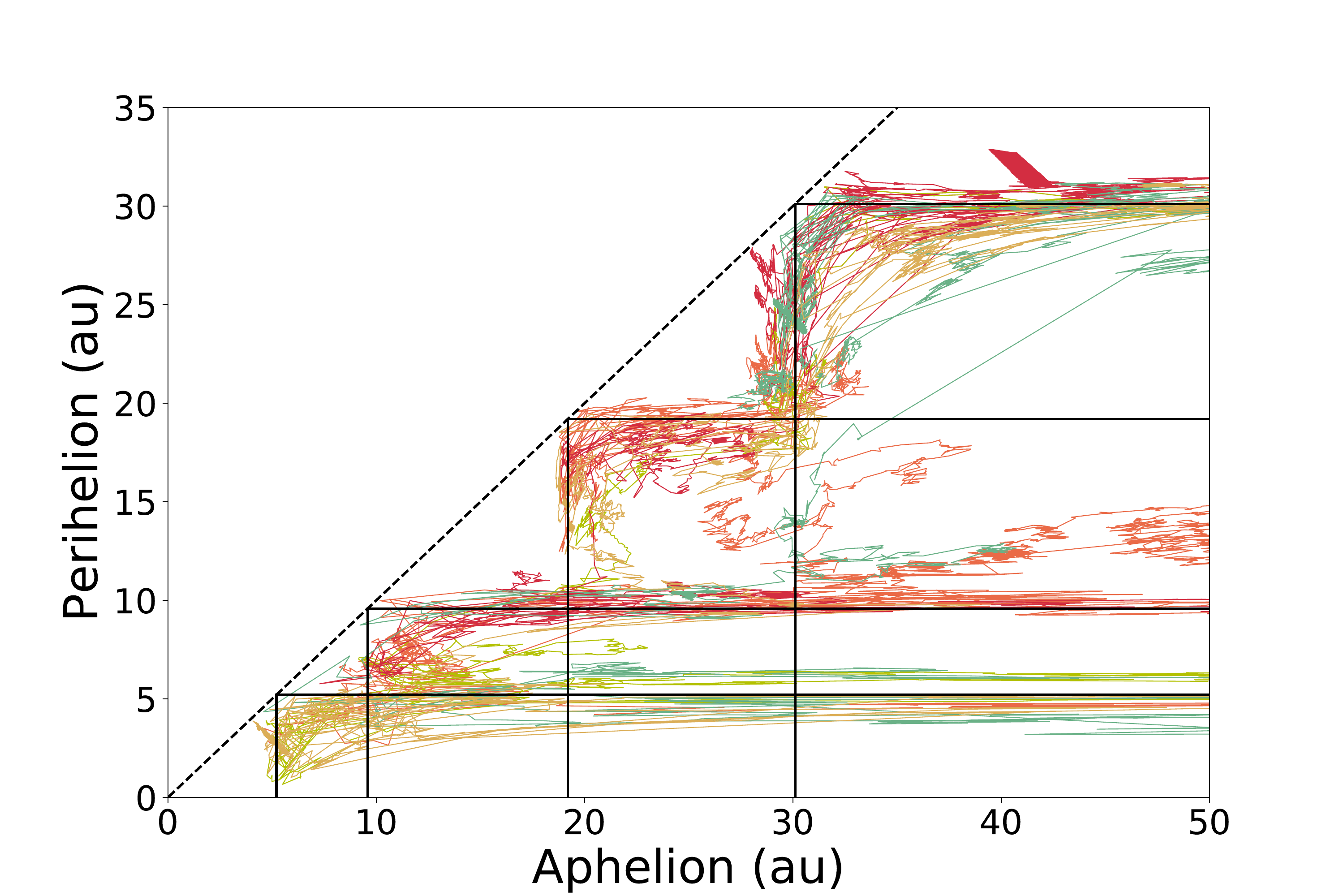}}
   \caption{The motion of particles in the $a$ {\it vs} $e$ plane. (a): The motion of over one hundred objects migrating from the Kuiper Belt to the JFC region showing their general dynamical evolution. (b): A few sample particles to show typical evolution more cleanly, instead with axes of perihelion and aphelion. The solid black lines are lines of constant perihelion or aphelion at the semimajor axes of the giant planets. \label{fig:centaur_walk}}
\end{figure*}

In a previous work, \citet{Bailey09} categorized the motion of the Centaurs using Hurst exponents (that is, they related the rms deviation of a particle's semimajor axis to time with $\sqrt{\langle a^2 \rangle} \sim t^H$, with the Hurst exponent $0.22 < H < 0.95$). They found that Centaurs undergo two broadly distinctive types of dynamical evolution, namely ``random walk'' behavior and resonance hopping. We observe these two types of motion in our simulations as well. Particles evolve under the influence of one or at most two of the giant planets as they migrate from the Kuiper Belt. While under the influence of a given planet, the particles' semimajor axes appear to vary randomly, but their perihelion or aphelion remains close to the orbit of the planet until passed off to another. We show this typical evolution of particles in Figure \ref{fig:centaur_walk}. As is evident, when particles are not in MMR, their motion in the plane is always loosely parallel to lines of constant perihelion or aphelion at the orbits of the planets (plotted as solid black lines). The only means by which particles can descend significantly far into the triangular spaces between planets is when they are trapped in resonance (a more in depth discussion of resonant behavior is presented in the analysis of Near Centaur dynamics). 

\subsection{Kuiper Belt Limits}
\label{sec:KBlim}

We can use our simulations to place broad limits on the population of small objects of the Kuiper belt, by comparing with the observable Jupiter Family Comets (OJFCs), which are defined as bodies with $2 < T_J < 3$ and $q < 2.5$ au \citep{Levison97,Rickman17}. The current number of known OJFCs is 355 according to the JPL's Small Body Database Search Engine. We note that the current population of OJFCs is very unlikely to be complete, and even if it was, we do not know the diameters of most of the objects. Nevertheless, we assume these objects are representative of the real OJFC population with $D > 2$ km for two reasons. First, it allows us to verify that our simulated OJFCs have similar orbital parameter distributions to a real population, as we do below. Secondly, the number $N_{\it obs} = 355$ is reasonably close to several independent estimates of the population: \citet{Rickman17} estimates it to be about 375-420, and \citet{Brasser15} find it to be on the order of 300, though they use $D > 2.3$ km. Using observed JFC sub-populations for estimating larger populations in this way is somewhat common in the literature, as is the case in \citet{DiSisto09} and \citet{Brasser15}, though they use substantially smaller, brighter, and closer populations that do not allow for the aforementioned validation. Therefore, it is reasonable to assume that this sample approximately represents the complete population of objects with $D > 2$ km \citep{DiSisto09,Brasser15}. We therefore assume the population of OJFCs with $D > 2$ km is complete and in steady-state.

Our original particles (Neptune crossers) are split into three different populations according to their source reservoir in the Kuiper belt (Classical, Resonant, and Scattering); as the intrinsic relative sizes between these three populations of the Kuiper belt are not known, we can only place upper limits on each population separately, by assuming for simplicity that each population contributes 100\% of the new comets.

\begin{figure}[!ht]
   \centering
   \includegraphics[width=\linewidth]{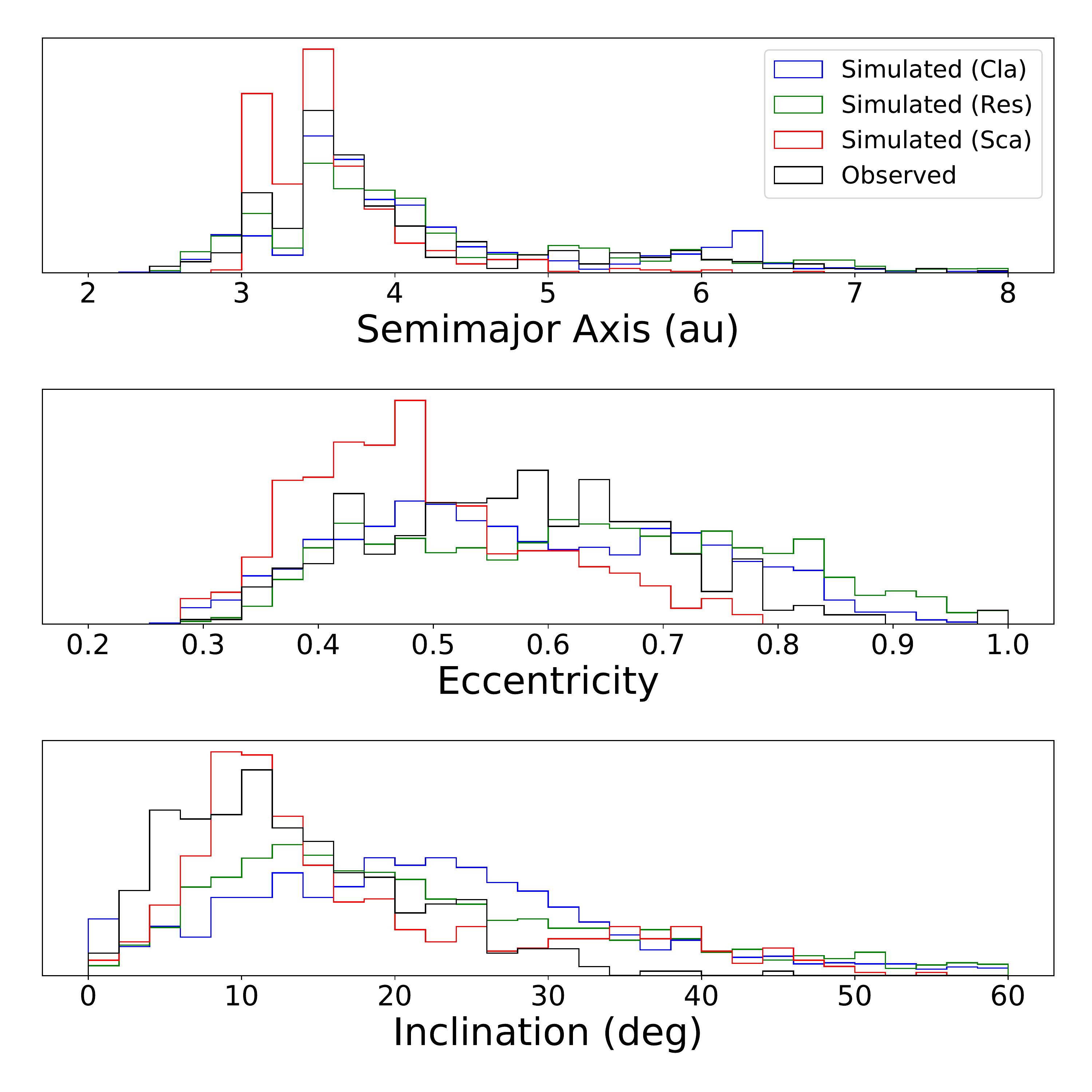}
   \caption{Normalized histograms of the real OJFCs and simulated OJFCs originating in the Classical, Resonant, and Scattering populations of the Kuiper belt. These plots are made by adding points to a histogram for every OJFC in our simulations every thousand years, and as such are very sensitive to single long-lived particles, especially when the total number of particles is low. This is the cause of the higher peaks in the Scattering semimajor axis distribution, which has two particles that spend several tens of thousands of years at the peaks among only 14 total OJFCs.} 
   \label{fig:JFChist}
\end{figure}

To ensure that the numerical results can confidently represent the population of OJFCs, we compare the distribution of OJFCs produced by our simulations with that of the real population of comets in Figure \ref{fig:JFChist}. The semimajor axis and eccentricity distributions are matched well by all our three populations, but the inclination distributions show a deficit of simulated particles at low inclinations. This seems to be a common issue from numerical simulations, also noted in previous studies; the inclination discrepancy is believed to be related to the physical evolution of cometary nuclei as they pass close to the Sun. In fact, it has been shown that a size-dependent modeling of the finite physical lifetime of comets can account for this problem. \citep{Rickman17,Nesvorny17,Sarid19}.

By means of our simulations, we can only directly limit the Neptune Crosser portion of the Kuiper belt; we can go one step further by using the fraction from the original populations in the Kuiper belt that became Neptune Crossers in the simulations of \citet{Munoz19}, from which we took our initial conditions for this work. The ratio of two populations $A$ and $B$ undergoing dynamic evolution at any given time is $\frac{N_A t_A}{N_B t_B}$, where $N_A$ is the number of objects that pass through the region and $t_A$ is the mean time they spend there. Our reservoirs were approximately static over the length of the simulation. Therefore, the size of the Neptune Crosser population $N_{\it cross}$ is:
\begin{equation}
N_{\it cross} = N_{\it obs}*\frac{N_{\it rsv}t_{\it sim}}{N_{\it OJFC}t_{\it OJFC}},
\end{equation}
where $N_{\it obs} = 355$ is the number of known OJFCs, $N_{\it rsv}$ is the size of the effective reservoir, $t_{\it sim}$ is the length of the simulation (20 Myr), $N_{\it OJFC}$ is the total number of particles that enter the OJFC region, and $t_{\it OJFC}$ is the average time spent in the OJFC region over the life of the particle. The limits are shown in Table \ref{table:KBlimits}. We also present the rate of new OJFCs, $N_{\it obs}/t_{\it OJFC}$. \citet{Rickman17} calculates this rate to be $(8.4 \pm 1.7) \times 10^{-3}$ per year, using a purely dynamical evolution and the same region definitions as ours, also assuming a steady state. Our rates agree well, with all three populations being within two standard deviations, though our uncertainties are high.

\begin{table*}[!ht]
\centering
\resizebox{\textwidth}{!}{\begin{tabular}{c c c c c c c c c}
\hline
Population & \thead{Reservoir\\ Size} & \thead{Total\\ OJFCs} & \thead{Time as \\OJFC (kyr)} & \thead{New OJFC Rate\\ ($10^{-3}$ yr$^{-1}$)} & \thead{Estimated\\ Crosser Population\\ ($\times10^{6}$ objects)} & \thead{Fraction \\of KB*} & \thead{KB Estimate\\ ($\times10^6$ objects)} & \thead{Observational\\ Upper Limit**\\ ($\times10^{6}$ objects)}\\
\hline
Classical & 4800 & 62 & 24.3$\pm$6.3 & 14.6$\pm$3.9 & 22.6$\pm$6.7 & 0.181 & 124.9$\pm$37.0 & 399 \\
\hline
Resonant & 4800 & 84 & 26.7$\pm$5.2 & 13.3$\pm$2.7 & 15.2$\pm$3.5 & 0.427 & 35.6$\pm$8.2 & 61 \\
\hline
Scattering & 4900 & 14 & 37.0$\pm$18.7 & 9.6$\pm$4.9 & 67.2$\pm$38.6 & 0.829 & 81.1$\pm$46.6 & 200 \\
\hline
\end{tabular}}
\caption{Estimates of the total Kuiper Belt population for three starting populations. Because the proportion of these populations is not known, each estimate assumes all known OJFCs come from that estimate's starting population. *Taken from \citet{Munoz19} **Estimated by \citet{Greenstreet19}
\label{table:KBlimits}}

\end{table*}

As the results of Table \ref{table:KBlimits} show, the number of required objects in any of the source populations of the Kuiper belt is well below an observational upper limit calculated independently by \citet{Greenstreet19}, based on the predictions of the number of craters on the surface of Arrokoth (formerly 2014~MU$_{69}$). We conclude that there exist more than enough objects in the reservoir trans-Neptunian region to account for the resupplying and maintenance of the population of JFCs in steady state, without the need to invoke new dynamical mechanisms for the delivery of objects than those already known \citep[see for example][]{Duncan95,Nesvorny17,Munoz19}.

\section{The Near Centaurs}

\subsection{29P \& 39P}

Comet 29P/Schwassmann-Wachmann 1 was discovered by a series of observations in November of 1927 during one of its outbursts. 
However, it has since been retroactively observed in a series of photographs from 1902. This gives 29P a significantly long history of observations for comparison with simulations; the Minor Planet Center (MPC) has orbital element data for it going back to 1908\footnote{ \url{https://minorplanetcenter.net/db_search/show_object?utf8=\%E2\%9C\%93&object_id=29P}}. This, combined with cloning, gives us a way to analyze the accuracy of our simulations on predicting the motion of real bodies. Notably, the observation record contains data from both before and after a short dynamic period around 1974 during which the eccentricity of 29P decreased from near 0.13 to the modern value of 0.04, among other changes.

\begin{figure*}[!ht]
   \centering
   \subfloat[][29P/Schwassmann-Wachmann\label{fig:29P}]
   {\includegraphics[width=0.475\textwidth]{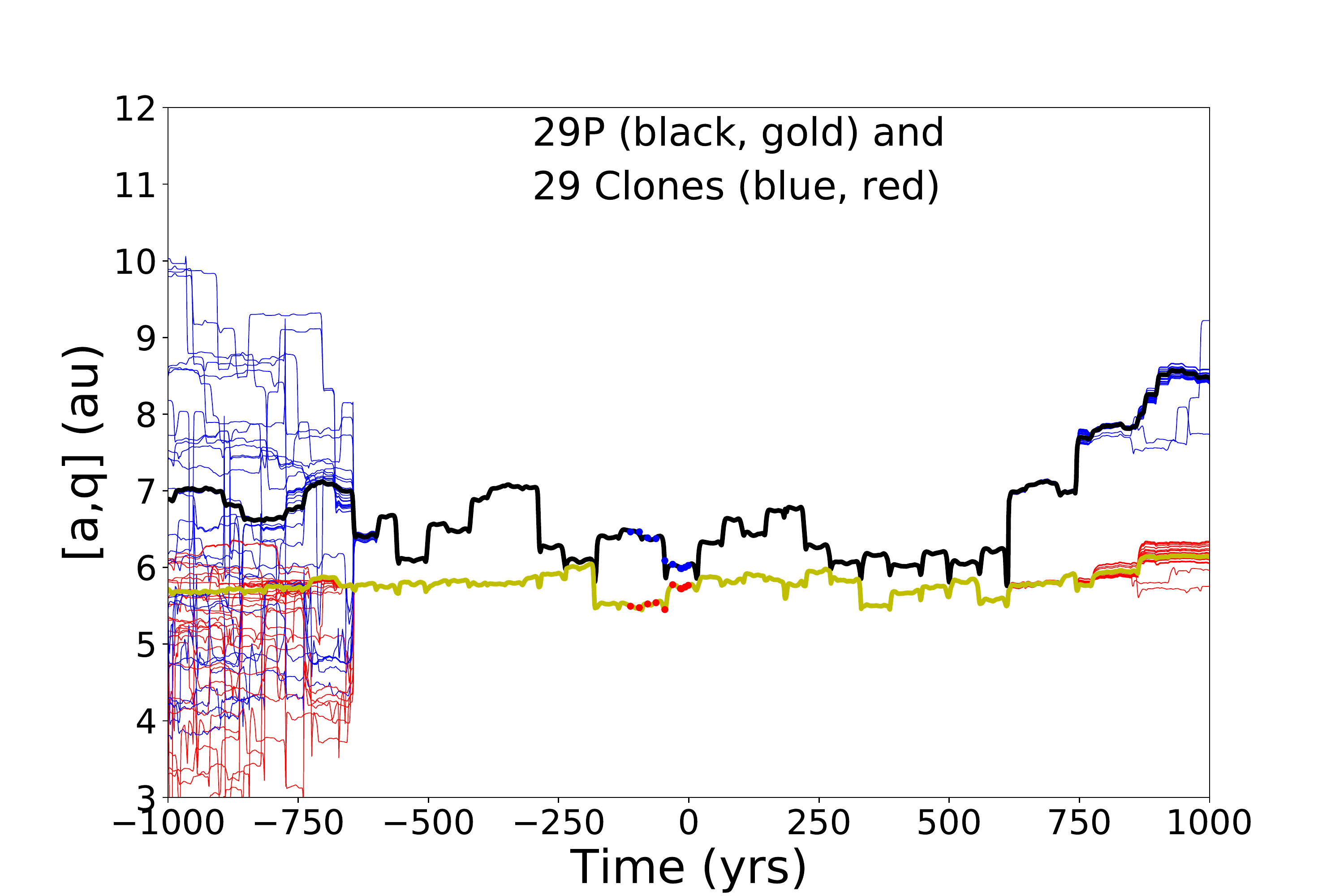}} 
   \subfloat[][39P/Oterma\label{fig:39P}] 
   {\includegraphics[width=0.475\textwidth]{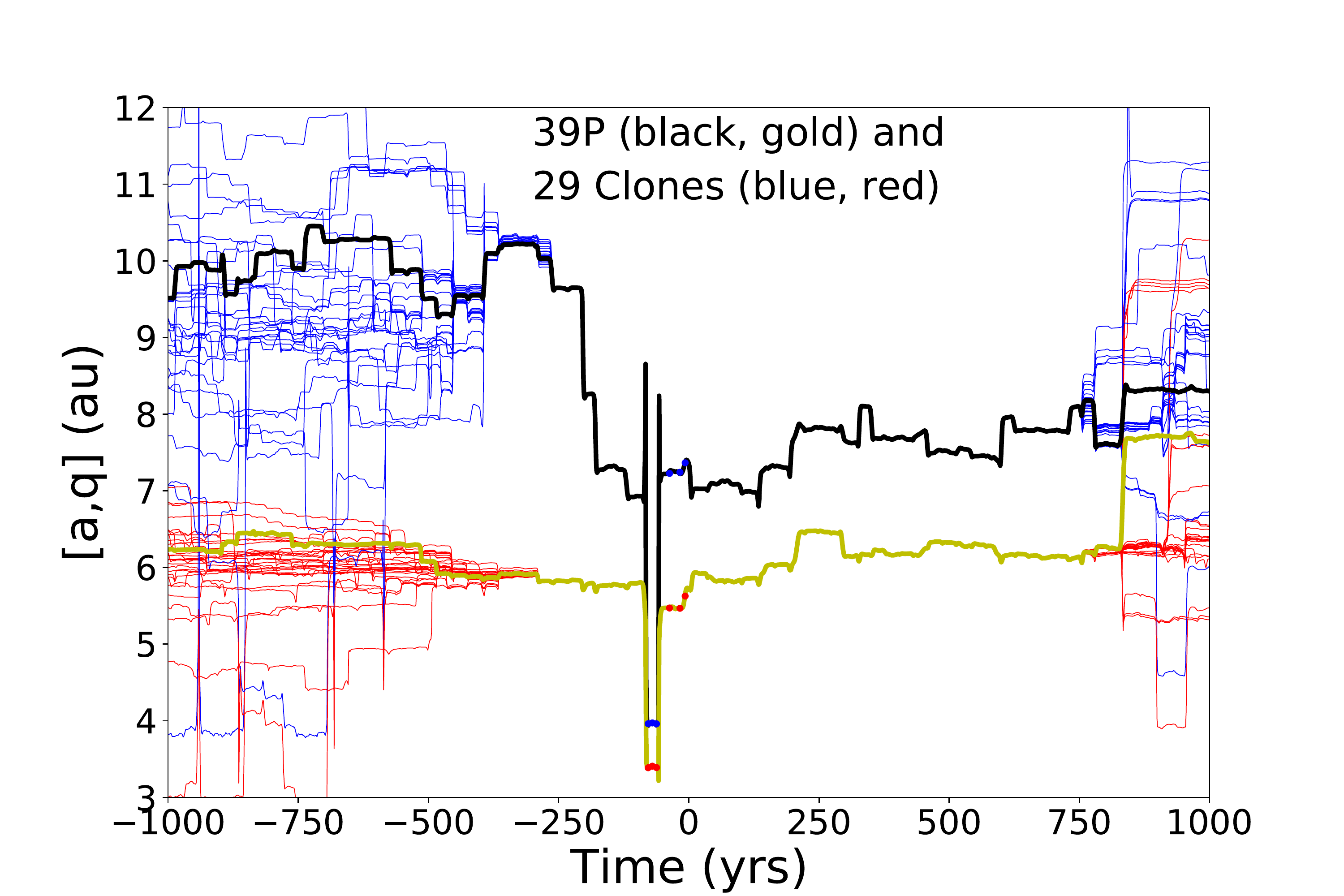}}
   \caption{Dynamic simulation of 29P and 39P and 29 clones one thousand years into the past and future. The red and blue points are the object's observed locations, beginning in 1908 for 29P and 1942 for 39P.}
\end{figure*}

The MPC gives the uncertainty on the perihelion distance of 29P to be about $5.2\times10^{-8}$ au. This is about a factor of two smaller than the $10^{-7}$ au variation on the Cartesian position of bodies when cloning in Section 2, so we used the same cloning technique for 29P. We integrated 29P and 29 clones forward and backward from the current date for one thousand years, with the other simulation parameters being identical to the simulations from Section 2. The semimajor axis and perihelion of 29P and its clones are plotted in Figure \ref{fig:29P}, and the same from the 10 MPC epochs of 29P since 1908 are plotted as points. Not only do the simulations agree very well with observations, but the particles are very consistent for $\pm 500$ years. 

Comet 39P/Oterma has a similarly long observational history, having been discovered in 1943; the MPC has orbital element data for it beginning in 1942\footnote{ \url{https://minorplanetcenter.net/db_search/show_object?utf8=\%E2\%9C\%93&object_id=39P}}. However, its history is significantly more variable than that of 29P. It was in an NCR orbit between Jupiter and Saturn for some time before it encountered Jupiter in 1937, which put it on an orbit completely interior to Jupiter's. It then became active, which allowed for its discovery several years later; however, this orbit was a 3:2 MMR and it encountered Jupiter once again in the 1960s, when it was put back onto a NCR orbit, though slightly more eccentric.

We performed an analogous simulation for 39P as that of 29P; the results are shown in Figure \ref{fig:39P}. The objects scatter much more quickly in the past than they do in that of 29P; however, the simulation does correctly calculate the two very close encounters with Jupiter discussed above. 

Though multiple very close encounters can make an object's position difficult to calculate, in general, our simulations seem to be very accurate on the order of hundreds of years. We therefore believe we can use short (hundreds of years) simulations to accurately assess the motion of the observed bodies in the NCR. 

\subsection{Dynamics of the Current Near Centaurs}
\label{sec:currentNC} 
 
\begin{figure}[!ht]
   \centering
   \includegraphics[width=\linewidth]{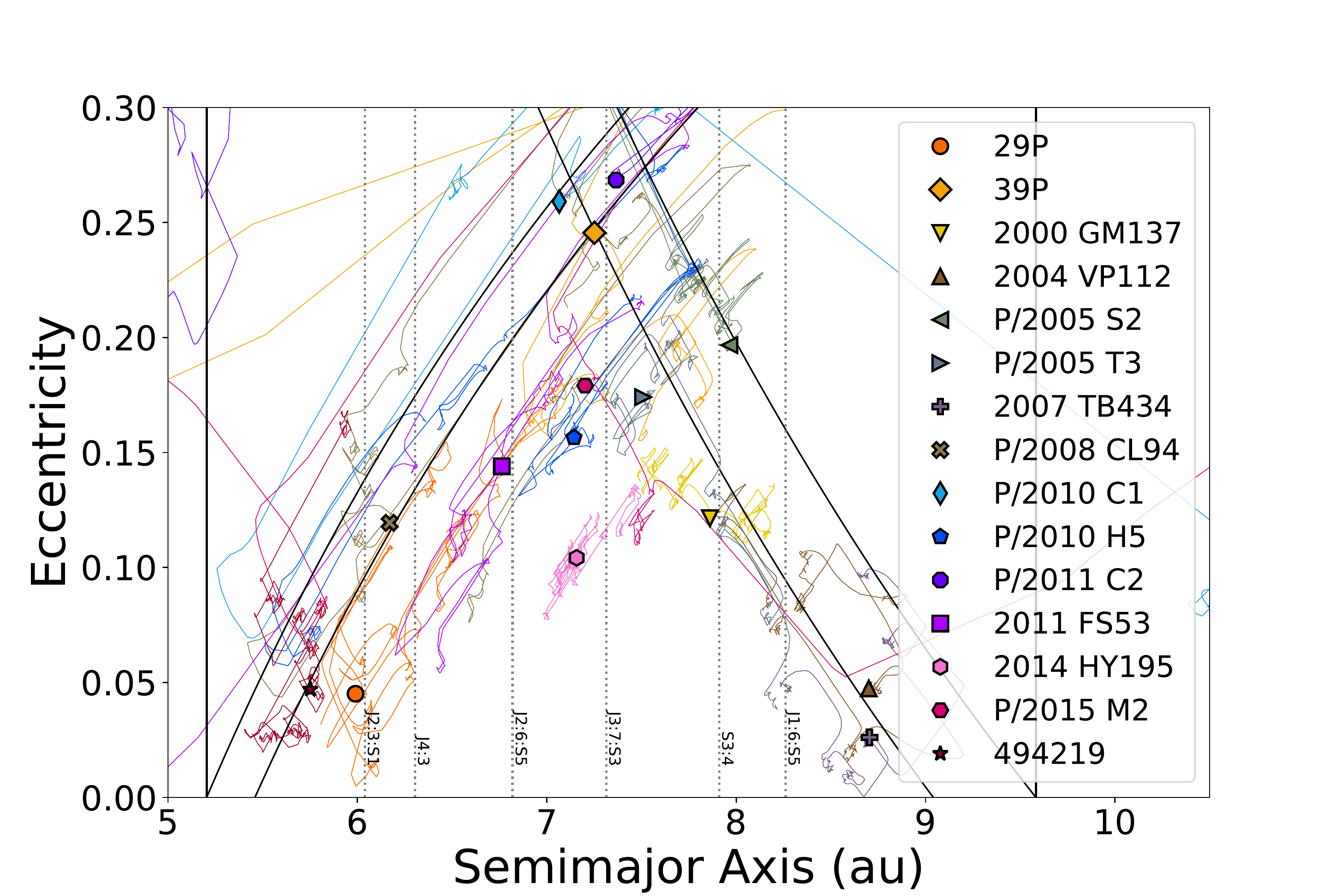}
   \caption{Dynamic simulation of the 15 current NCR objects to $\pm 500$ years, plotted in the a-e plane.   \label{fig:NCpath}}

\end{figure}

Based on the above discussion, we integrated the fifteen current NCR objects to $\pm 500$ years from the current date. Their paths in the a-e plane are plotted in Figure \ref{fig:NCpath}. Of the fifteen current, only eight were in the region 500 years ago, and only 11 will be 500 years in the future.    

We categorize the fifteen Near Centaurs into four dynamical classes. The first is those that are evolving primarily under the influence of Jupiter; in Figure \ref{fig:NCpath} these bodies move parallel to Jupiter's perihelion line. 29P/Schwassman-Wachmann, P/2008 CL94 (Lemmon), P/2010 H5 (Scotti), 2011 FS53, and 494219 are all in this class. The second is those that are evolving primarily under the influence of Saturn; they move loosely parallel to Saturn's aphelion line and tend to remain in the NCR longer than those under Jupiter's influence. This class includes 2000 GM137, 2004 VP112, P/2005 S2 (Skiff), P/2005 T3 (Read), and 2007 TB434. The third class is objects that are influenced strongly by both Jupiter and Saturn. Their orbital evolution is highly chaotic, and they are unlikely to remain Near Centaurs longer than a few hundred years. 39P/Oterma, P/2010 C1 (Scotti), P/2011 C2 (Gibbs), and P/2015 M2 (PANSTARRS) are in this class.

\begin{figure*}[!ht]
   \centering
   \subfloat[][P/2005 S2 (Skiff) \label{fig:P/2005_S2}]
   {\includegraphics[width=0.475\textwidth]{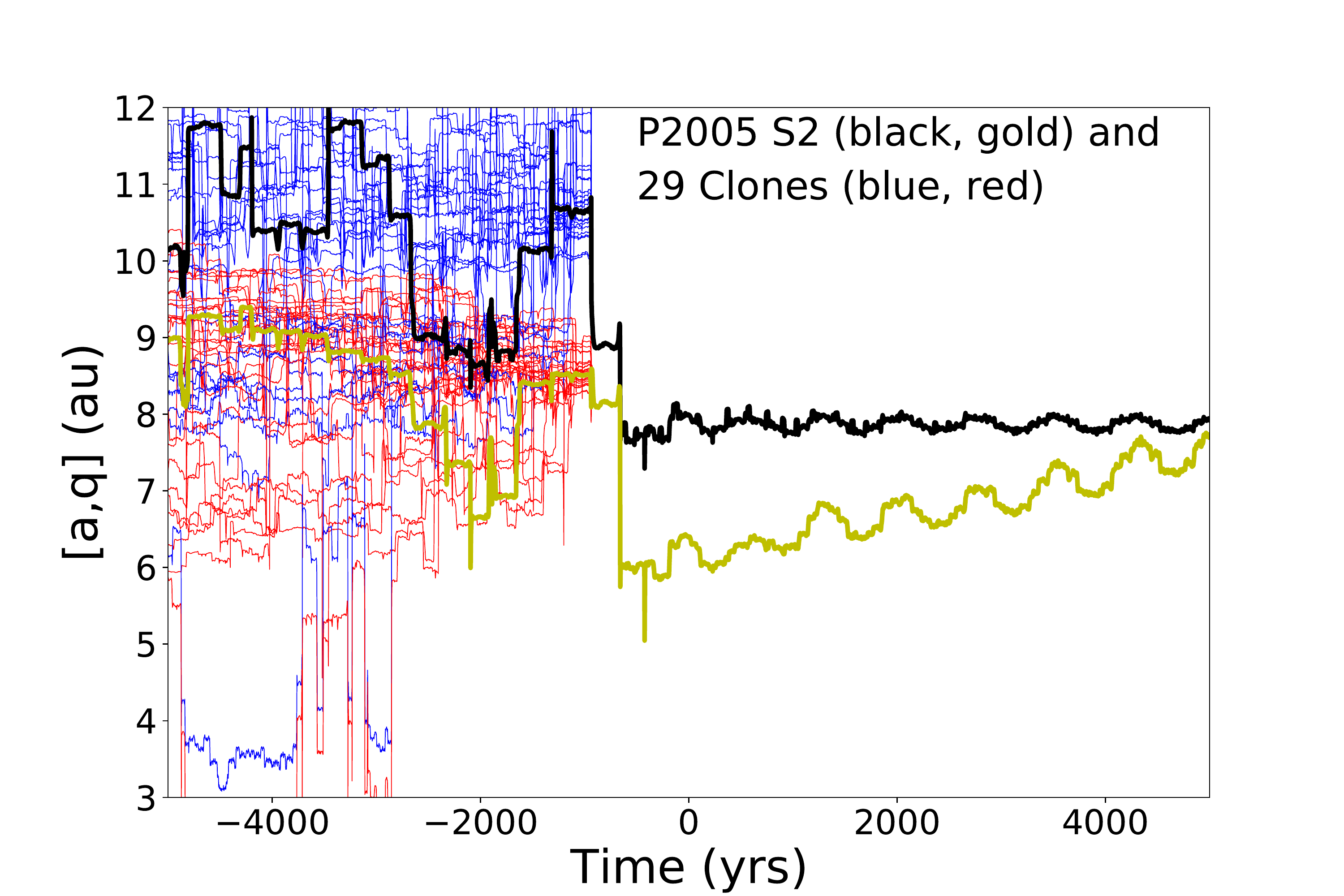}}
   \subfloat[][2014 HY195 \label{fig:2014_HY195}] 
   {\includegraphics[width=0.475\textwidth]{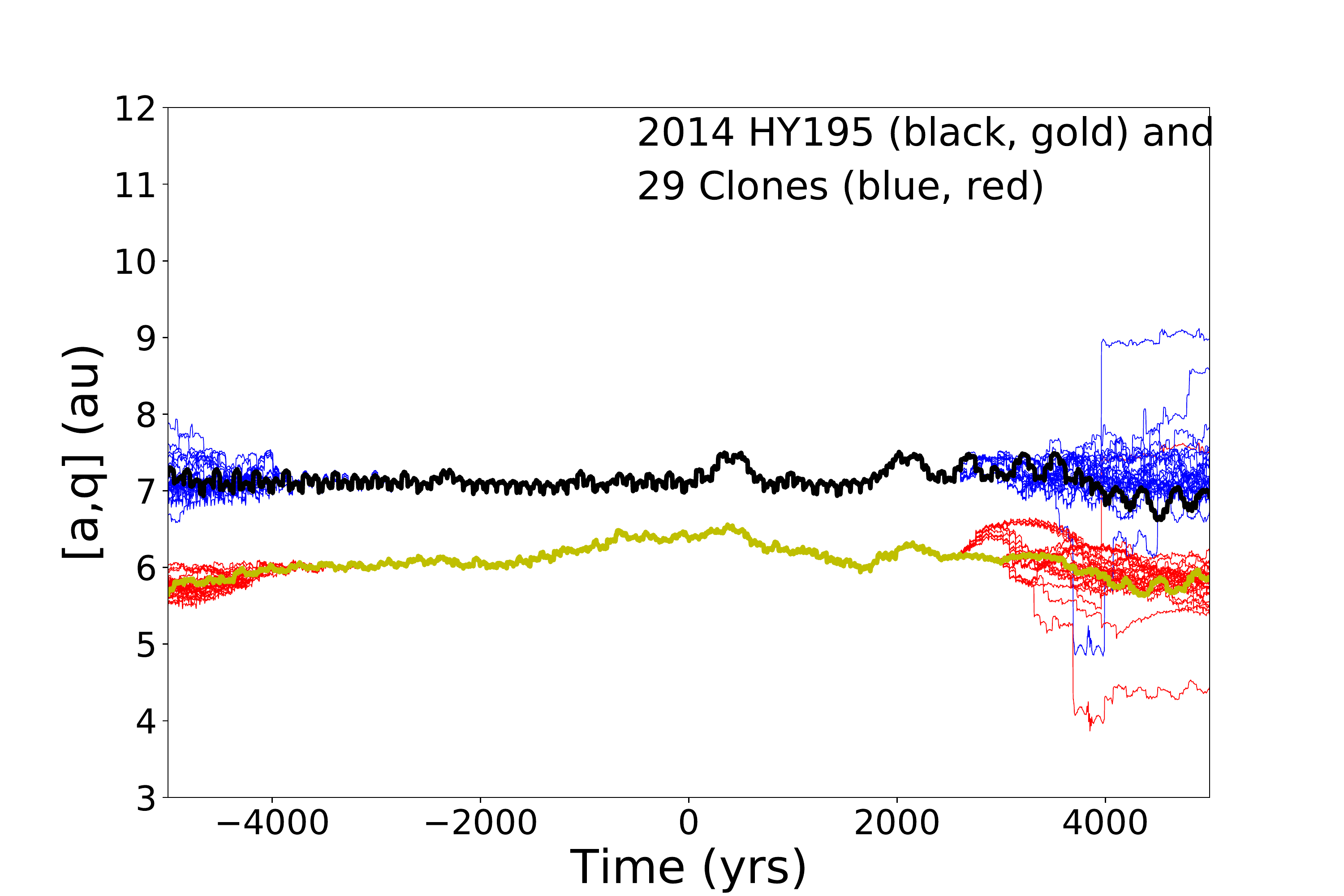}}
   \caption{Dynamic simulation of resonant bodies P/2005 S2 (Skiff) and 2014 HY195 and 29 clones five thousand years into the past and future.}
   \label{fig:resonant_NCs}
\end{figure*}

The last class deserves longer discussion. From Figure \ref{fig:NCpath}, it is clear that 2010 HY195 is moving in a resonant way (right in the middle of the NCR), though it is not obvious which resonance it is caught in. Further simulation to $\pm 5000$ years of each of the observed NC objects (and their 29 clones), revealed that 2010 HY$_{195}$ is indeed caught in resonance and will stay so for at least 1000 years, though it appears to briefly hop to another resonance and then hop back in the next several centuries. More interestingly, these simulations also revealed that P/2005 S2 (Skiff) is currently or will soon become caught what seems to be Saturn's 3:4 resonance and remain there for at least 5000 years into the future. P/2005 S2 (Skiff) is classified as under the influence of Saturn because it is not clear if it is currently in resonance, but it is clear that its past dynamics are primarily determined by the influence of Saturn. The evolution of both of these bodies to $\pm 5000$ years from the present is presented in Figure \ref{fig:resonant_NCs}.

\subsection{Simulated Near Centaurs}

\begin{figure}[!ht]
   \centering
   \includegraphics[width=\linewidth]{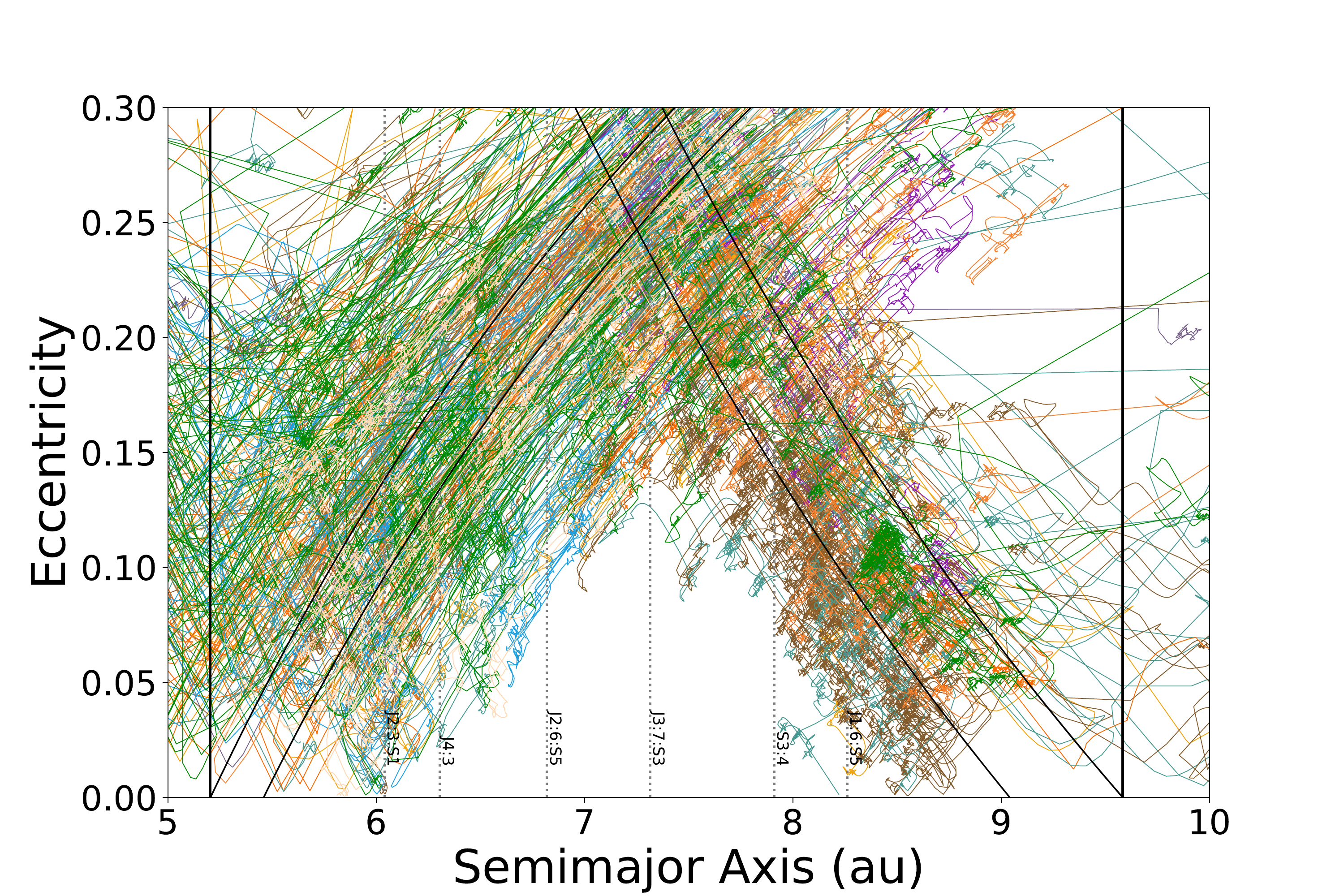} 
   \caption{Paths of non-resonant simulated particles in the NCR.}
   \label{fig:Non_res}
\end{figure}

If a particle entered the NCR at any of the thousand-year checkpoints of our 20 Myr simulations, its time in the region was re-simulated and its elements recorded at 0.1 year intervals for more detailed analysis. The paths of a collection of particles which never became trapped in MMR in the a-e plane are presented in Figure \ref{fig:Non_res}. The dynamics of the non-resonant classes discussed in Section \ref{sec:currentNC} are clearly replicated by the simulations. Just as with the real NCR objects, the simulated particles close to Saturn's orbit evolved much more slowly than those near Jupiter's, and particles were most chaotic when near both. It seems that objects in this region exhibit behavior primarily influenced by Jupiter when they have perihelion approximately $q < 6$ au and with Saturn when they have aphelion approximately $Q > 8.4$ au.

Particles only descended below this triangle in $a$-$e$ phase space when in resonant motion. 15.6\% of particles that passed through the NCR stayed in a single resonance for more than five kiloyears, which was determined by tracking how long particles' semimajor axes stayed within $0.3$ au of any of the MMR locations between Jupiter and Saturn. 3.0\% of particles that passed through the region showed resonant behavior for more than 50 kyr (across one or more separate resonances). Of the many resonances between Jupiter and Saturn, particles were most frequently caught in the four non-pure 3BR discussed above (see Fig. \ref{fig:FAmap}), Jupiter's 4:3, and Saturn's 3:4.

\begin{figure*}[!ht]
\subfloat[]{
\begin{minipage}[8cm]{.24\textwidth}
\centering
\includegraphics[width=3cm,height=2cm]{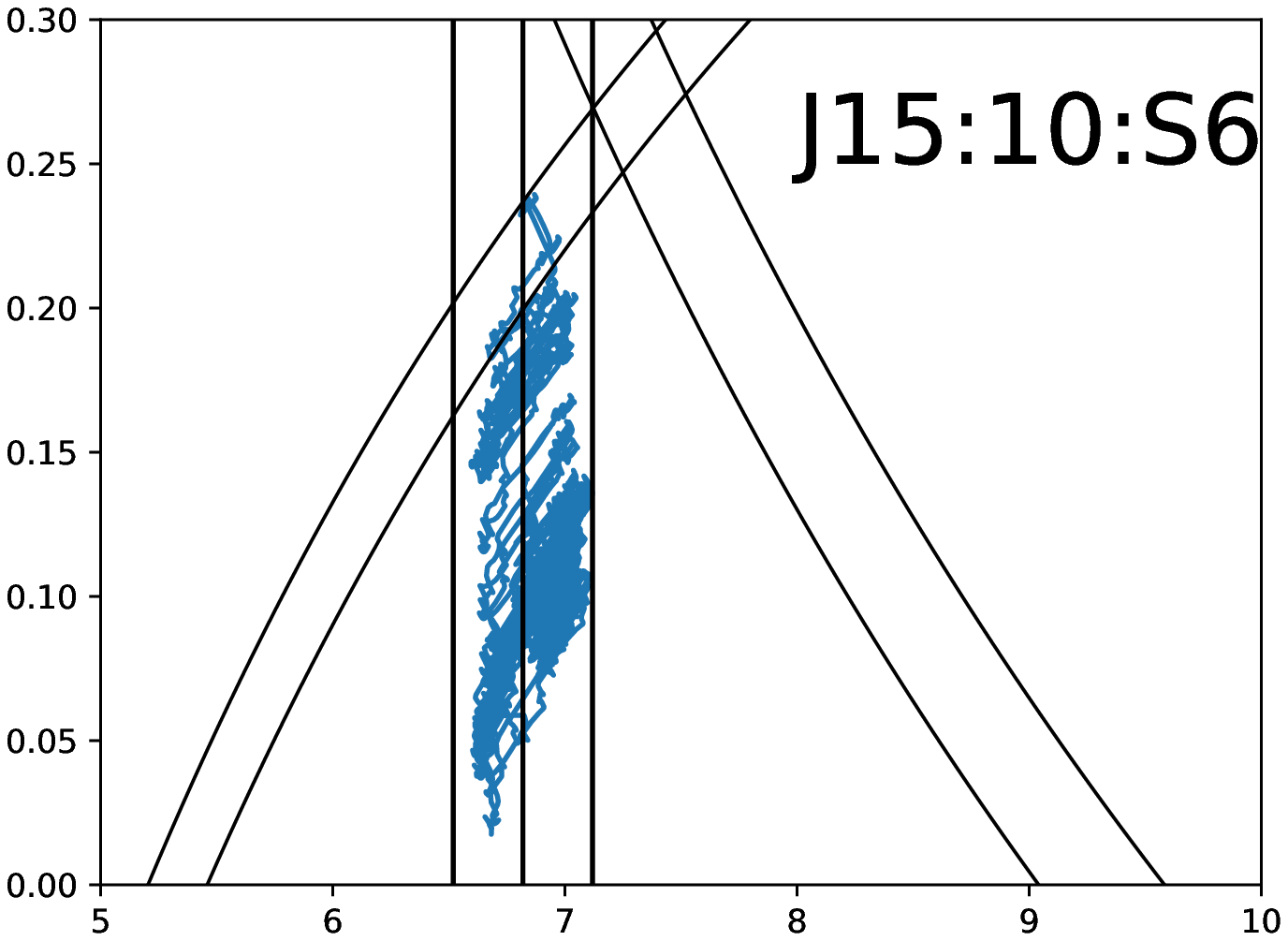}
\vfill
\includegraphics[width=3cm,height=2cm]{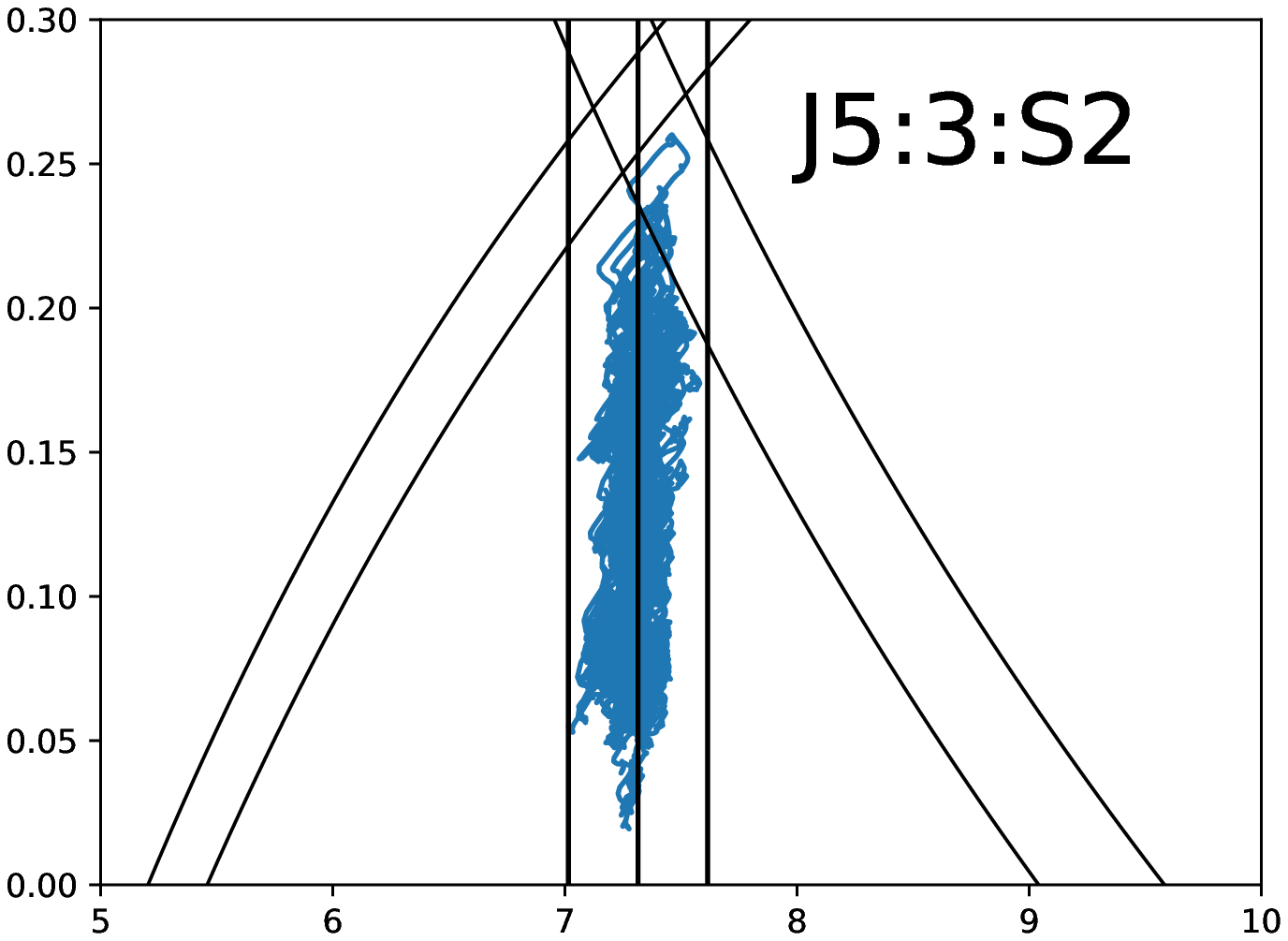} 
\vfill
\includegraphics[width=3cm,height=2cm]{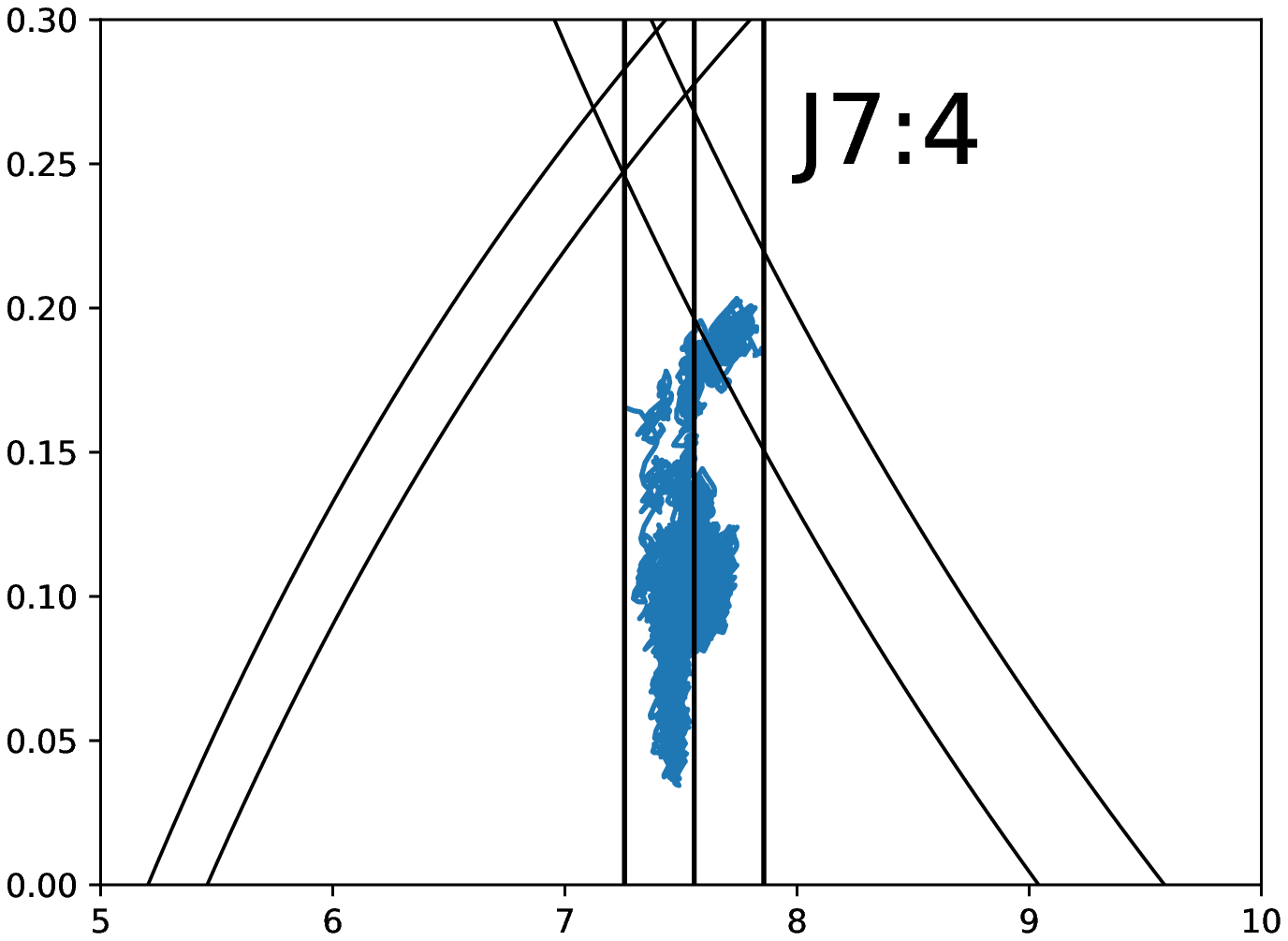}
\end{minipage}}
\subfloat[]{\begin{minipage}[c][6.5cm]{.74\textwidth}
\includegraphics[width=10cm,height=6.5cm]{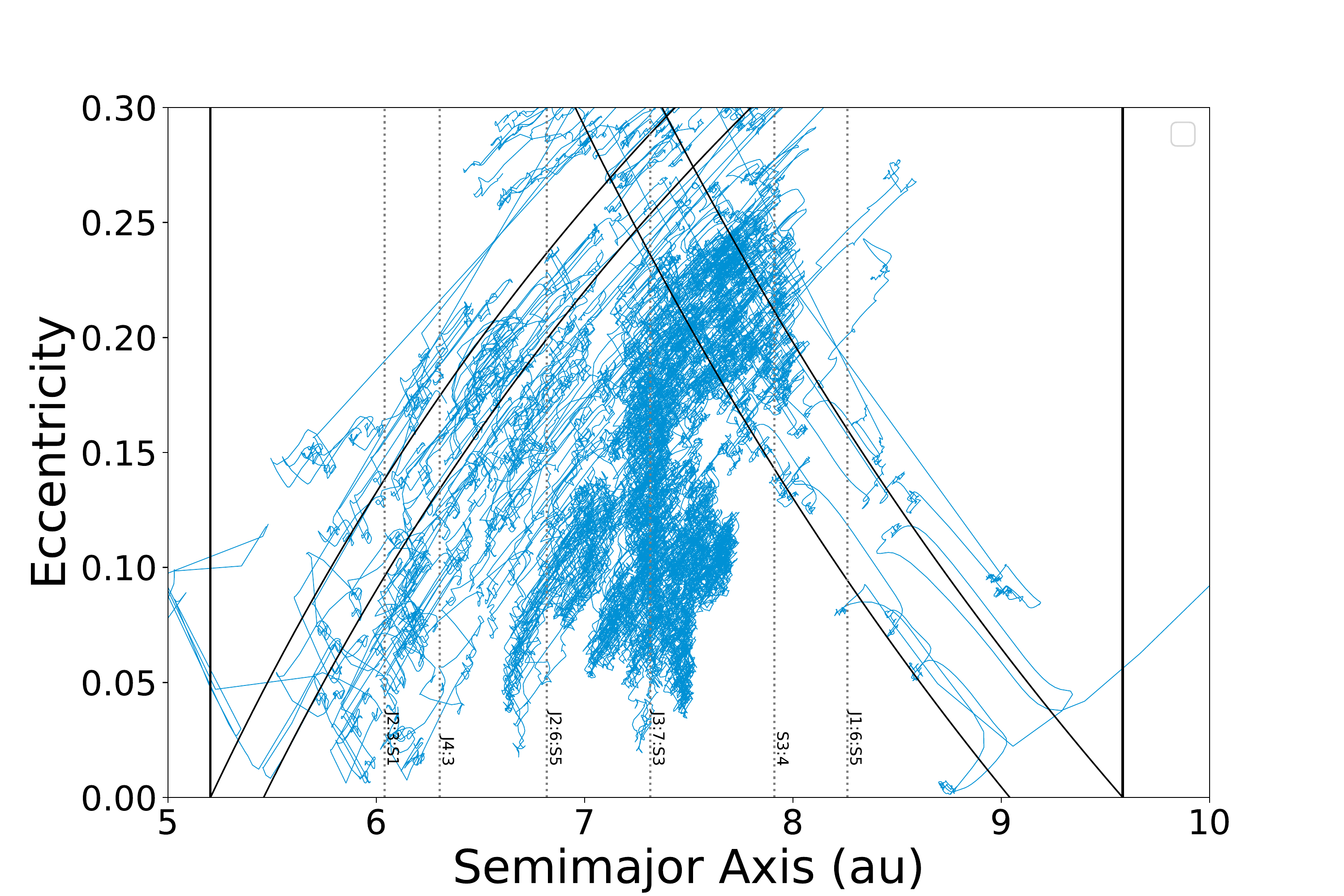}
\end{minipage}}
\caption{Three individual resonances (a) which compose parts of a particle's full path through the NCR (b). The length of time spent in each resonance is 12 kyr, 33 kyr, and 27 kyr from top to bottom. The particle occupied a single resonance for more than 5 kyr 11 times and spent a total of almost 80 kyr in resonance.\label{fig:res_hop}}
\end{figure*}

It is much of the time difficult to concisely distinguish resonances from one another. Particles frequently hopped between resonances and even spent time influenced by two or more resonances at once. Figure \ref{fig:res_hop} is an example of a particle that spent much of its time in the NCR hopping between resonances. The longest time a particle spent in a single resonance was 103.6 kyr in the Jupiter-Saturn 3BR J2:6:S5, expressed by the combination $2n_J-6n_P+5n_S\approx0$. 

\subsection{Near Centaur Population Estimate}
\label{sec:NCRpop}
 
As in Section \ref{sec:KBlim}, we can estimate the size of the population of the NCR using the known population of OJFCs. The population of real NC objects, $N_{\it rNC}$, in relation to the known OJFCs, $N_{\it obs}$, is:
\begin{equation}
N_{\it rNC} = N_{\it obs}*\frac{N_{\it NC}t_{\it NC}}{N_{\it OJFC}t_{\it OJFC}},
\end{equation}
where $N_{\it NC}$ is the total number of particles that became NC objects at any time and $t_{\it NC}$ is the mean length of time a particle spent in the region. In both the OJFC region and the NCR, particles have long forgotten their Kuiper Belt origin, so the three groups can be combined to increase the statistical significance of the calculation. The estimates are presented in Table \ref{tab:NClimits}. The total number of NC objects at any given time is about $\frac{3}{4}$ the number of OJFCs, meaning the current 15 known NC objects represent roughly 6\% of the expected population.

\begin{table*}[!ht]
\centering
\begin{tabular}{||c|c c c c c c||} 
\hline
Population & Reservoir & \multicolumn{2}{c}{Total Number} & \multicolumn{2}{c}{Time in Region (kyr)} & Near Centaur Limit \\ [0.5ex]
 & & OJFCs & NCs & OJFCs & NCs & \\
\hline\hline
Classical & 4800 & 62 & 102 & 24.3$\pm$6.3 & 10.9$\pm$1.7 & 261$\pm$90 \\
\hline
Resonant & 4800 & 84 & 133 & 26.7$\pm$5.2 & 13.3$\pm$2.0 & 279$\pm$79 \\
\hline
Scattering & 4900 & 14 & 19 & 37.0$\pm$18.7 & 14.4$\pm$4.2 & 187$\pm$127 \\
\hline
Combined & 14500 & 160 & 254 & 26.6$\pm$4.0 & 12.4$\pm$1.3 & 262$\pm$55 \\
\hline
\end{tabular}
\caption{ An estimate of the total number of Near Centaurs. }
\label{tab:NClimits}
\end{table*}

\section{Discussion}

\subsection{Origin and Evolution of Centaurs}

In recent years, thanks to the vast increment in the inventory of observed populations of small bodies in the solar system, several works have focused on the detailed, quantitative characterization of the relationship between JFCs, their outer Solar System reservoirs, and the Centaur region between them.

Numerical simulations are the main tool for studies aiming to determine the link between Centaurs and JFCs. Like in the present analysis, most works include both long-term (for a large-scale dynamical study) and short-term integrations (to follow the evolution of single objects). Recent works focusing on the origin, evolution, and distribution of Centaurs include for example \citet{Fernandez18} and \citet{DiSisto20}. \citeauthor{Fernandez18} explore the differences between the dynamical evolution of active and inactive Centaurs, using both short and long-term integrations; by their part, aiming to update a previous study \citep{DiSisto07} regarding the origin and distribution of Centaurs, by including the new set of observed objects, \citet{DiSisto20} re-simulated the evolution of their increased sample, now splitting the classical Centaur region into one defined only by semimajor axis limits, as well as a Giant Planet-Crossing (GPC) region defined by perihelion limits. In particular, they study the Scattered Disk as the source of these two populations, finding it to be the largest contributor to both of them.

\subsection{Relevance of 29P for Centaurs and JFCs}

\begin{figure}[!ht]
   \centering
   \includegraphics[width=\linewidth]{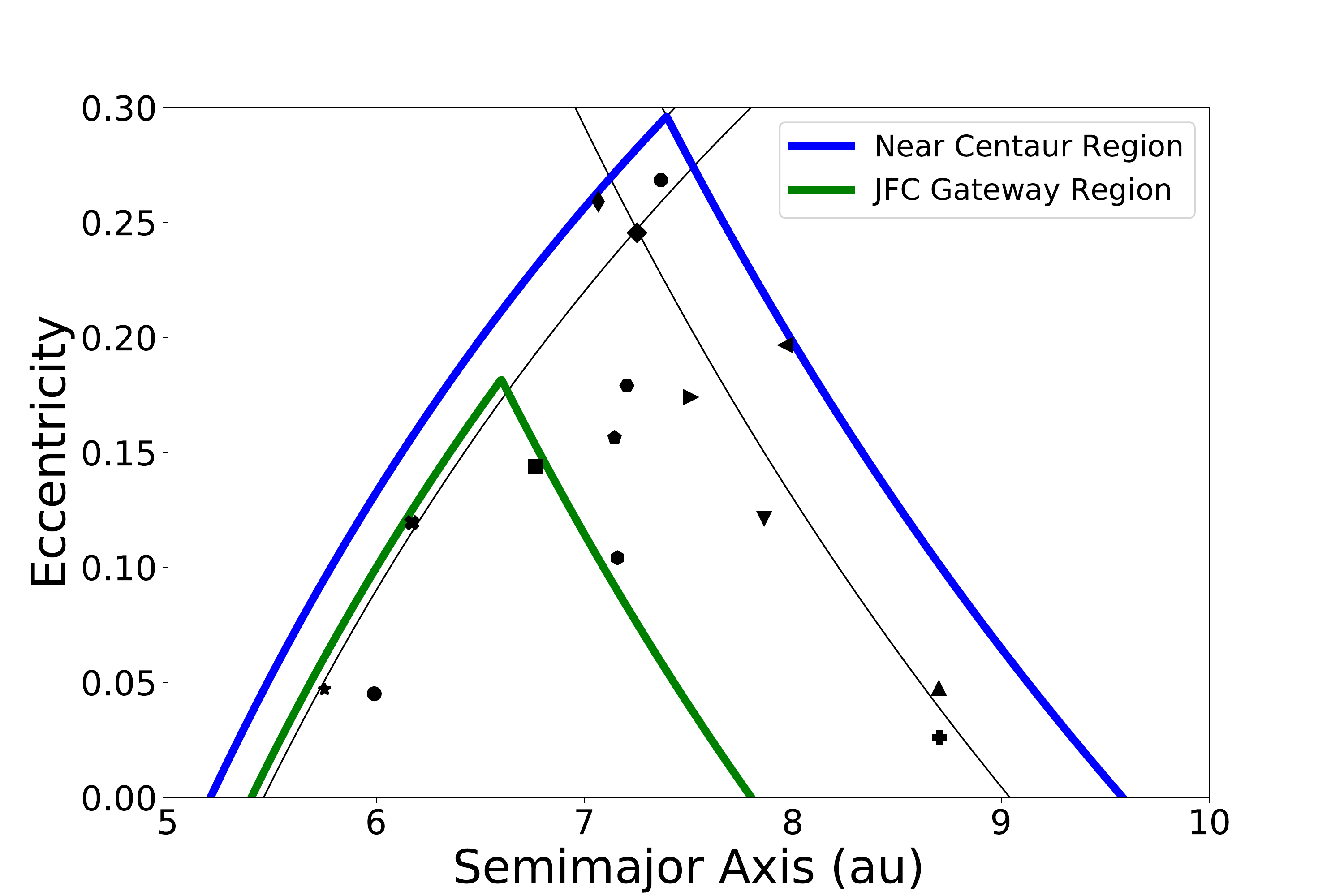} 
   \caption{Region definitions for the Near Centaurs and the JFC Gateway. Included are points for the 15 current NCR objects. The thin black lines show the constant perihelion at the aphelion of Jupiter and the line of constant aphelion at the perihelion of Saturn.}
   \label{fig:gNCs}
\end{figure}

Other recent studies have been drawn by the interest in the unique orbit and physical properties of 29P/Schwassman-Wachmann 1. Along this line, \citet{Sarid19} analyze a region similar to, though smaller than, our Near Centaur region, which they call the JFC Gateway, drawn by 29P, its most prominent occupant. The JFC Gateway region is somewhat similar to our NCR, though its limits, $q > 5.4$ au and $Q < 7.8$ au, are much tighter than the NCR; the definitions of the JFC Gateway region and our NCR are shown in Figure \ref{fig:gNCs} for comparison (also note that \citeauthor{Sarid19} define the JFCs and Centaurs differently than in this work). \citeauthor{Sarid19} numerically explore the dynamical evolution of objects starting from the trans-Neptunian region, until they become JFCs. In particular, they focus on the dynamics of Centaurs as they pass through the Gateway region before becoming JFCs. However, they do not analyze resonant behavior which, as shown in this work, is very important for the long-term dynamics and stabilization of trapped objects. 

In contrast with our work, \citeauthor{Sarid19} included in their simulations a fading law to model physical evolution of comets \citep[from][]{Brasser15}, in addition to using a larger number of particles than we do. Despite such differences, our simulations and results generally agree. We present a comparison of some statistics in Table \ref{tab:Sarid}, showing some of the results obtained by using the \citeauthor{Sarid19} definition of the Gateway region within our simulations. 

29P is also specifically discussed in \citet{Nesvorny19}, which uses the discoveries of the OSSOS survey \citep{Bannister18} to model the dynamics and orbital distribution of Centaurs, as well as to set strict constrains on the size of the population. They find the lifetime for objects in orbits similar to 29P to be about 38 kyr; our value of 12.4 kyr for the NCR is similar in magnitude, but our value is decreased by the inclusion within the NCR of many more chaotic regions.

\subsection{Size Population of Centaurs}

The subject of the size of the Centaur population is complicated by conflicting definitions of the Centaurs themselves. Thus, it becomes difficult to simply compare estimates of the Centaur population available throughout the literature and validate results. In Table \ref{tab:CentaurPop} we present a comparison between several estimates from recent works, as well as estimates based on our simulations when using the Centaur definitions given in those works. Our estimates are consistent with, though lower than, those of \citet{Nesvorny19} and \citet{DiSisto20}. It is surprising that our simulations predict the same amount of Centaurs (to two significant figures) for three distinct definitions of the Centaurs, especially when the definitions of this work and that of \citet{DiSisto20} are so different. This occurs because the additional $a$-$e$ phase space of the \citet{DiSisto20} definition where $q<a_J$, a large though quickly evolving area, contributes nearly identically to the Centaur population estimate as the additional space in our definition where $30<a<a_N$, which is small but densely populated. 

The fact that three distinct Centaur definitions \citep[those of][and ours]{Nesvorny19,DiSisto20} lead to similar population estimates suggests that such slight differences in the particular definitions do not have as large of an impact as might be expected; still, such differences complicate the direct comparison among published results. We argue that, to avoid this issue, a standardization of the Centaur definition is pertinently on time, given the growing amount of data available for comparison both between works, as well as between simulations and observations. 

Finally, we note that our simulations predict a factor of about 50 fewer Centaurs than \citet{Sarid19}, when using their definition. Their calculation uses a power law Centaur size distribution drawn from cratering on the Pluto-Charon system \citep{Singer19}, the size distribution of observed JFCs \citep{Snodgrass11}, and the fragmentation of C/1999 S4 \citep[LINEAR][]{Makinen01}. This estimate would be an upper limit due to its use of cratering, but it is still high, especially for a very restrictive definition of the Centaurs. Our simulations support a lower population, as do the two results in the previous paragraphs, but a deeper investigation into this issue would be productive.  

\subsection{Non-gravitational Effects}

In this work, we do not model the sublimation of cometary nuclei in order to simplify our simulations and analysis. However, it is well established that sublimation will influence comet dynamics, depending on the size of the object, by creating non-gravitational forces due to outgassing. Though such forces are likely very small for an object as large as 29P, it is conceivable that they could be significant for less massive bodies. However, it does not seem likely that outgassing will strongly affect the general dynamical behavior we observe in the NCR; though small forces could push an object away from a resonance, there are other resonances nearby that can sustain resonant behavior (that is, it might cause a body to `hop' in an indistinguishable way to the object in Figure \ref{fig:res_hop} does). Furthermore, the primary source of outgassing for the JFCs, water ice, does not substantially sublimate at distances beyond the orbit of Jupiter, so the dynamical effect of sublimation is limited to other sources, such as CO and CO$_2$ \citep{Jewitt09}.

In addition to outgassing, the continuous activity causes some comets to be exhausted or even destroyed after many close passes to the Sun. As we discuss in Section \ref{sec:KBlim}, activity plays a crucial role in shaping the inclination distribution of the JFCs. \citet{Sarid19} confirm this by implementing a fading law from \citet{Brasser15} in their simulations, finding that without fading, comets spend too much time at low perihelion, allowing their inclinations to become higher than is otherwise possible with a shorter lifetime. Furthermore, the existence or non-existence of activity in Centaurs can also be related to their histories and future evolution. As found by \citet{Fernandez18}, active Centaurs are far more likely to become JFCs than inactive Centaurs, which have a much wider range of final $T_J$. They argue that, though both active and inactive Centaurs likely originate from the trans-Neptunian region, inactive Centaurs have different histories, where they may be flung outward and influenced by galactic tides before entering the Centaur region.

\subsection{Resonances in the NCR}

Resonance is a common feature in analyses that focus on JFCs and Centaurs; for example, \citet{Fernandez18} discuss the impact of resonances interior to Jupiter on the dynamics of JFCs. On the other hand, although much work has been done regarding the effect of 3BRs in the Asteroid belt \citep[see for example][]{Nesvorny98,Smirnov13}, no other works study the influence of resonances between Jupiter and Saturn. We show strong evidence that resonant behavior is common in the NCR, opening an avenue for a longer stability of objects in this region than is otherwise expected. This is particularly interesting because previous studies \citep[e.g.][]{Robutel01} did not find any stable locations in this zone of the solar system. In this work we identified resonant behaviour in non-pure 3BR, which result from a chain of 2 two body MMR between a particle and either Jupiter or Saturn. The main locations where particles can survive on a long term basis are associated with chains formed by overlapping first order MMRs with Jupiter and Saturn. In particular we found two 3BRs of order zero, expressed by the combinations $2n_J-3n_P+n_S\approx0$ at 6.04 au (a Laplace resonance) and $n_J-6n_P+5n_S\approx0$ at 8.26 au, as well as two 3BRs of order one, expressed by the combinations $2n_J-6n_P+5n_S\approx0$ at 6.82 and $3n_J-7n_P+3n_S\approx0$ at 7.32 au. We believe the possible long-term stability on such locations merits further investigation, currently outside the scope of this paper.

\begin{table*}[!ht]
\centering
\resizebox{\textwidth}{!}{\begin{tabular}{c c c c c c c}
\hline
Study & \thead{Fraction of\\ JFCs in Gateway} & \thead{Fraction of Gateway\\ Particles in JFC Region} & \thead{Fraction of Centaurs\\ entering Gateway } & \thead{Mean Time in\\ Gateway (yrs)} & \thead{Median Time\\ in Gateway (yrs)} & \thead{Fraction Gateway Particles\\ Reaching $q < 3$ au}\\
\hline
\citet{Sarid19} & 0.72 & 0.77 & 0.21 & 8000 & 1750 & 0.49 \\
This Work & 0.52 & 0.90 & 0.114 & 5960 & 2000 & 0.48 \\
\hline
\end{tabular}}
\caption{A comparison between predictions for particles in the NCR and the Gateway region. All values in this table use the region definitions from \citet{Sarid19}. \label{tab:Sarid}}
\end{table*}

\begin{table*}[!ht]
\centering
\resizebox{\textwidth}{!}{\begin{tabular}{c c c c}
\hline
Study & Centaur Region Definition & Population Estimate & Our Estimate \\
\hline
\citet{Sarid19} & $q > 5.2; \ Q < 30.1$ & $6.5 \times 10^6$ & $1.4 \times 10^5$\\
\citet{Nesvorny19} & $a < a_N; \ q > 7.5$ & $6 \times 10^5$ & $3.6 \times 10^5$\\
\citet{DiSisto20} & $5.2 < a < 30$ & $~10^6$ & $3.6 \times 10^5$ \\
This work & $a_J < (a,q) < a_N$ & - & $3.6 \times 10^5$ \\
\hline
\end{tabular}}
\caption{A comparison between estimations of the Centaur population size with their respective definitions. All estimates listed here are for the population of Centaurs with diameter $D > 2$ km. \label{tab:CentaurPop}}
\end{table*}

\section{Conclusions}

In this work, we present the results of high-precision numerical simulations of test particles evolving under the influence of the Sun and the planets. Our initial conditions are given by a population of crossers derived by previous long-term simulations \citep[the population of crossers after 1 Gyr of evolution in][]{Munoz19}, i.e. particles currently crossing Neptune's region of influence. We followed particles at high cadence as they moved into the inner solar system, with a focus on their behavior while in the NCR between Jupiter and Saturn.

We confirmed the dynamical behavior noticed in previous works. The ``random walk'' behavior exhibited by particles in the Centaur region is shown to be related to objects evolving with constant perihelion or aphelion under the influence of one or two of the giant planets. On the other hand, resonant evolution inside any of the MMRs with the giant planets, in some cases can lead to a lowering of the eccentricity, thus increasing the residency time of centaurs in the giant planet region. We found the mean time spent by Centaurs in the planetary region to be 2.70 Myr, with a median of 1.7 Myr. 

The mean lifetime for objects in the NCR is roughly 10-15 kyr, with about 13 new NCR objects per kyr. Comparison using the observed OJFC population constrains the size of the Kuiper Belt to within observational limits, with large uncertainties. 

Short-term simulation of the observed Near Centaurs divides their dynamics into groups being influenced by Jupiter, Saturn, both planets, or exhibiting resonant behavior; P/2005 S2 (Skiff) and 2014 HY195 are shown to be in resonance for several millennia around the present. Simulations show resonant Near Centaur behavior is common and that particles can stay in resonance for up to 100 kyr, suggesting a mechanism for comparatively long-term stability in the region. 

Comparison with the observed OJFC population suggests that about 250 objects with $D > 2$ km are currently located in the NCR. 

We discuss the place of the present analysis among several recent works and compare our results. This paper is one of several that study the dynamics of Centaurs as a population and 29P/Schwassmann-Wachmann 1 as a notable occupant. Many model the sublimation of cometary nuclei and its effects on dynamical evolution, but none study resonance as we do. Our estimate of the size of the Centaur population with $D>2$ km agrees well with several previous studies, though the matter is complicated by differences in definition used throughout the literature. We stress the need for a standardization of a Centaur definition in order to confidently compare the results among different studies.

\section*{Acknowledgements}

We acknowledge Julio Fernandez and an anonymous referee for insightful reviews that help to improve the quality of this paper. We also thank A. P. Granados and M. Alexandersen for useful discussions. AR acknowledges support from the ASIAA Summer Student Program.

\bibliography{ms}

\end{document}